\documentclass[sigconf]{acmart}
\RequirePackage{etex}

\usepackage{tikz}
\usepackage{amsmath,amsfonts}
\usepackage{relsize}
\usepackage[utf8]{inputenc}
\usepackage{dblfloatfix}
\usepackage{arcs}
\usepackage{color}
\usepackage{graphicx}
\usepackage{latexsym}
\usepackage{psfrag}
\usepackage{soul}
\usepackage{color, colortbl}
\usepackage{algorithm}
\usepackage[vlined,ruled,linesnumbered,algo2e]{algorithm2e}
\usepackage[font=small,skip=10pt]{caption}
\usepackage[tight]{subfigure}
\usepackage{float}
\usepackage{pifont}
\usetikzlibrary{arrows,shapes}
\usepackage{multirow}
\usepackage{url}
\usepackage{xspace}
\usepackage{hyperref}
\usepackage{mdframed}
\usepackage{enumitem}
\usepackage{appendix}

\newcommand{\eg}{{\it e.g., }}
\newcommand{\ie}{{\it i.e., }}
\newcommand{\sol}{{Fides}\xspace}

\newcommand{\tea}{{service}\xspace}
\newcommand{\stu}{{verification}\xspace}
\newcommand{\ddf}{{GDTL}\xspace}

\newcommand{\pecserver}{{server}\xspace}

\newcommand{\servpro}{{service provider}\xspace}

\usepackage{amsmath,scalerel}
\DeclareMathOperator*{\Bigcdot}{\scalerel*{\cdot}{\bigodot}}
\floatname{algorithm}{Algorithm}
\SetKwComment{Comment}{$\quad  \triangleright$\ }{}
\SetCommentSty{itshape}

\makeatletter
\newcommand{\nosemic}{\renewcommand{\@endalgocfline}{\relax}}
\newcommand{\dosemic}{\renewcommand{\@endalgocfline}{\algocf@endline}}
\let\oldnl\nl
\newcommand{\nonl}{\renewcommand{\nl}{\let\nl\oldnl}}
\makeatother

\begin{document}
\fancyhead{}
\title{A Generative Framework for Low-Cost Result Validation of Machine Learning-as-a-Service Inference}
\authors{
    \author{Abhinav Kumar}
    \affiliation{%
      \institution{Saint Louis University}
      \city{St. Louis}
      \country{U.S.}}
    \email{abhinav.kumar@slu.edu}

    \author{Miguel A. G. Aguilera}
    \affiliation{%
      \institution{New Mexico State University}
      \city{Las Cruces}
      \country{U.S.}}
    \email{guirao@nmsu.edu}
    
    \author{Reza Tourani}
    \affiliation{%
      \institution{Saint Louis University}
      \city{St. Louis}
      \country{U.S.}}
    \email{reza.tourani@slu.edu}
    
    \author{Satyajayant Misra}
    \affiliation{%
      \institution{New Mexico State University}
      \city{Las Cruces}
      \country{U.S.}}
    \email{misra@nmsu.edu}
    \renewcommand{\shortauthors}{A. Kumar \kETAL}
}
\acmYear{2024}\copyrightyear{2024}
\acmConference[ASIA CCS '24]{ACM Asia Conference on Computer and Communications Security}{July 1--5, 2024}{Singapore, Singapore}
\acmBooktitle{ACM Asia Conference on Computer and Communications Security (ASIA CCS '24), July 1--5, 2024, Singapore, Singapore}
\acmDOI{10.1145/3634737.3657015}
\acmISBN{979-8-4007-0482-6/24/07}
%
\begin{abstract}
The growing popularity of Machine Learning (ML) has led to its deployment in various sensitive domains, which has resulted in significant research focused on ML security and privacy. However, in some applications, such as Augmented/Virtual Reality, integrity verification of the outsourced ML tasks is more critical--a facet that has not received much attention. 
Existing solutions, such as multi-party computation and proof-based systems, impose significant computation overhead, which makes them unfit for real-time applications. 
We propose {\em\sol}, a novel framework for real-time integrity validation of ML-as-a-Service (MLaaS) inference. \sol features a novel and efficient distillation technique--Greedy Distillation Transfer Learning--that dynamically distills and fine-tunes a space and compute-efficient \stu model for verifying the corresponding \tea model while running inside a trusted execution environment.
\sol features a client-side attack detection model that uses statistical analysis and divergence measurements to identify, with a high likelihood, if the \tea model is under attack. \sol also offers a re-classification functionality that predicts the original class whenever an attack is identified. We devised a generative adversarial network framework for training the attack detection and re-classification models.
The evaluation shows that \sol achieves an accuracy of up to 98\% for attack detection and 94\% for re-classification. 
\end{abstract}

%
%

\maketitle
%
\section{Introduction}
\label{sec:intro}
The increasing popularity of machine learning (ML) applications and the plethora of data generated by smart and connected devices, such as smartphones, Internet of Things devices, and video cameras, has led to the development of sophisticated and complex ML applications, such as autonomous driving and cognitive assistance~\cite{ZhaZhaShi18}. However, the challenges of processing such a high volume of data to support real-time applications have led to the development of several edge-computing solutions for autonomous driving~\cite{LiuLiuTan19}, healthcare~\cite{ray2019edge}, and Augmented/Virtual Reality applications~\cite{siriwardhana2021survey}.

\begin{figure}[t]
\centering
    \includegraphics[width=\columnwidth]{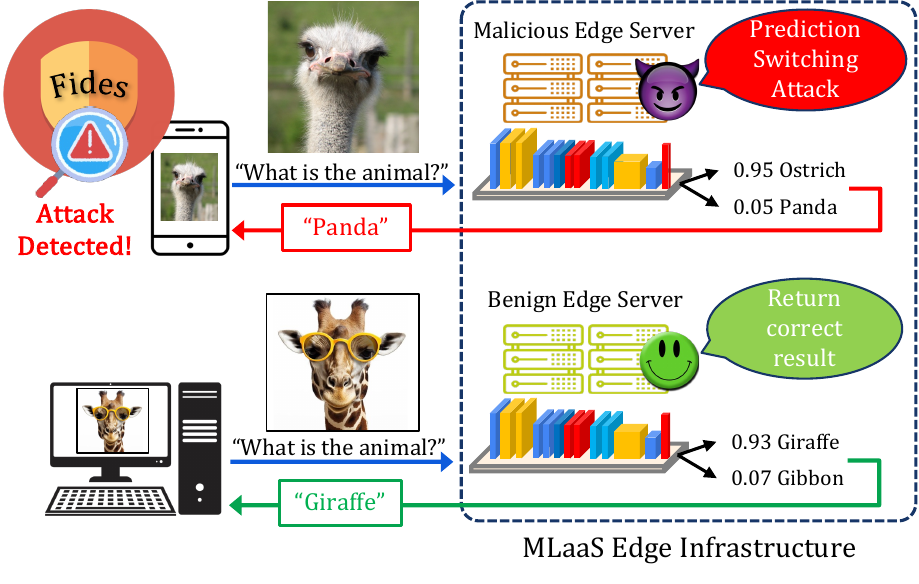}
    \vspace{-0.3in}
\caption{We consider the integrity verification of Machine Learning-as-a-Service inference, where clients send their data to the edge servers for ML inference tasks. In our proposed framework, \sol, we aim to detect any malicious misclassification caused by a malicious edge server when running the clients' inference tasks.}
\label{fig:div_intro}
\vspace{-0.2in}
\end{figure}

This motivates inference-based ML-as-a-Service (\ie MLaaS), in which trusted ML providers deploy their services on third-party edge servers as query-based APIs. These APIs allow the clients to share their data with the edge servers, which are running ML applications, to provide the processed result.
However, the edge computing trust model differs from the Cloud, primarily due to its distributed and multi-stakeholder nature~\cite{TouSriMis20}. Given that any entity may host a third-party server, outsourcing ML applications to the edge ecosystem raises potential privacy and security issues (Figure~\ref{fig:div_intro}), including model/data inference and Trojan attacks~\cite{LiuLiZho18,NasShoHou19,TanDuLiu20,LiuDinSha21,zhang2021trojaning,cheng2021deep,TraShoSan22}. 
Two of the most pertinent attacks against the integrity of MLaaS are Trojan attacks~\cite{zhang2021trojaning,cheng2021deep} and adversarial example~\cite{papernot2016transferability,tao2018attacks}. In Trojan attacks, the attacker partially modifies the neural network using a poisoned dataset to include a set of samples with a pattern trigger, aiming to cause misclassifications when the model observes the pattern trigger. Backdoor attacks can be devised either during model training or on a pre-trained model~\cite{TanDuLiu20,zhang2021trojaning,cheng2021deep,YaoLiZhen19}. An adversarial attack (\ie adversarial example) exploits the inherent vulnerabilities of neural networks for post-training misclassifications. These attacks are input-specific in that each input data should be individually perturbed to cause a misclassification.

\begin{figure}[t]
\centering
    \includegraphics[width=\columnwidth]{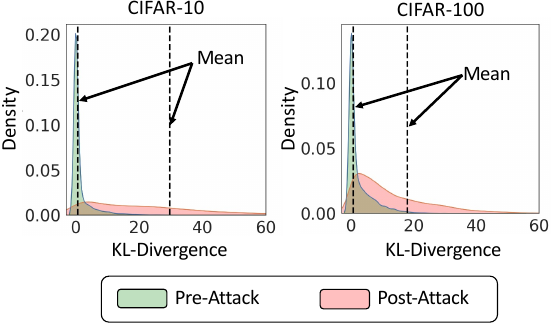}
    \vspace{-0.2in}
    \caption{The KL divergence density distribution between two models ({\bf ResNet50} and {\bf ResNet152}) before (green) and after (red) attack on one of the models (ResNet152). Both models are trained on Cifar10 and Cifar100 datasets. The KL divergence between the benign models' posterior vectors belongs to a distribution with low mean and variance (green distribution). Performing a prediction-switching attack against one of the models (ResNet152) leads to a significant increase in the distribution's mean and variance (red distribution). We use this identifiable behavior in designing \sol. (distributions' tails are truncated for better visualization).}
    \label{fig:intro_2}
\vspace{-0.3in}
\end{figure}

The existing techniques for preserving the integrity of outsourced computation use cryptosystems, such as multi-party computation~\cite{huang2022cheetah,sotthiwat2021partially,yuan2021practical,knott2021crypten}, proof-based systems~\cite{ghodsi2017safetynets,lee2020vcnn,madi2020computing}, and homomorphic encryption~\cite{niu2020toward,xu2020secure,madi2020computing,natarajan2021chex}. A few initiatives have built frameworks using a Trusted Execution Environment (TEE) to offload parts of the ML applications~\cite{TraBon18,grover2018privado}.
Despite their merits, these solutions are often computationally expensive, impose significant latency overhead, or do not scale with the complexity of the evolving ML models. Thus, making them impractical for real-time and latency-sensitive applications. 
Some solutions utilize multiple models in the form of redundant computing or majority voting~\cite{AbdInjSha18,hashemi2021darknight}. Redundant computing techniques, however, require higher degrees of redundancy to achieve stricter integrity guarantees, which leads to significantly higher computational overhead. Moreover, redundant computing techniques are built on the presumption that the only relevant part of a redundant model's output is the class with the highest probability (predicted class), which is untrue. \emph{The model's posterior vector provides information about the learned knowledge representation through the probabilities assigned to the incorrect classes.} This insight is used in knowledge distillation for training higher accuracy and fidelity compressed versions of cumbersome models~\cite{HinVinDea15}.
We believe the same insight can be used to study the impact of an adversary on two models trained on identical distributions. Figure~\ref{fig:intro_2} shows the density distribution of Kullback–Leibler (KL) divergence between two models (ResNet50 and ResNet152) and the drastic change in the distribution when an adversary manipulates the prediction generated by one of the models (ResNet152). We will analyze the impact of an adversary on the divergence trends (Section~\ref{sec:observation}) to create a result validation framework.\\

\noindent
\textbf{Our Framework:}
To cope with the attacks that target MLaaS inference integrity, we propose \sol. 
In a nutshell, \sol validates the integrity of MLaaS inference by securely running a special-purpose {\it verification model}. This model is constructed through knowledge distillation, guaranteeing that the knowledge representation of the \stu model closely approximates that of the \tea model.
%
\sol comprises two primary components. 
The first one is a resource-efficient model distillation technique--Greedy Distillation Transfer Learning (\ddf)--which generates a customized \stu model for a given \tea model. \ddf distills the knowledge of the \tea model into the \stu model by incrementally unfreezing and fine-tuning the last layers of the model. This leads to only a few layers being distilled at a time, which makes it suitable for resource-constraint and TEE-based private training paradigms~\cite{hashemi2021darknight,MoHadKat21}
The second component is a novel generative framework for training a client-side attack detection model that uses posterior vectors of the \tea and \stu models, alongside their divergences, on a given input to detect a potential attack. Our framework also trains an attack re-classification model, aiming to predict the correct result of the outsourced inference task by learning a typical adversary's goal and behavior in the system under attack.

We evaluate \sol' performance using three datasets, including CIFAR-10, CIFAR-100, and ImageNet, across three neural network architectures and also assess its computational overhead on multiple constrained devices. We compare \sol with Slalom~\cite{TraBon18} and Chiron~\cite{HunTylSho18}, and show that \sol achieves $4.8 \times$--$26.4 \times$ and $1.7 \times$--$25.7 \times$ speed-up to Slalom and Chiron, respectively. \\
%

%
%

The novel {\bf contributions} of \sol are as follows: 
%

    \noindent {\bf (i)} We propose \sol--a framework for result validation of MLaaS inference using an efficient verification model running on a TEE to corroborate the results of the \tea model. This makes \sol a perfect candidate for real-time applications, such as autonomous driving, and medical imaging. 
    
    \noindent {\bf (ii)} In \sol, we propose a greedy distillation transfer learning (\ddf) technique for training the verification model, aiming to reduce the distillation's overhead by incrementally unfreezing layers of the verification model for fine-tuning to the desired accuracy. 
    
    \noindent {\bf (iii)} We introduce two client-side shallow neural networks for detecting and resolving adversarial attacks, with negligible computation overhead. We propose a generative framework for effective training of these models. 
    
    \noindent {\bf (iv)} We systematically assess \sol using major benchmark datasets of varying complexity and number of detection classes on three architecturally different deep neural networks: namely ResNet~\cite{resnet}, DenseNet~\cite{densenet}, and EfficientNet~\cite{efficientnet}. Our results show that \sol outperforms existing solutions in terms of computation complexity while achieving high attack detection and remediation accuracy.
%
%

We share the observation that motivated our design in Section~\ref{sec:observation}. Section~\ref{sec:background} presents the background. In Section~\ref{sec:models}, we elaborate on system model, threat model and assumptions, and the detail of the attack modeling. Section~\ref{sec:design}, we present the detail of our design.
Section~\ref{sec:evalution} presents our experimental insights. Section~\ref{sec:related} includes the related work. Section~\ref{sec:conclusion} shares our conclusion.

\section{Observation}
\label{sec:observation}
%
In this section, we will discuss our findings regarding the disparities between two models, which were trained on datasets with similar distributions, where one is subjected to an attack that causes unintended mispredictions. The insights drawn from these observations, coupled with the evident trend we have identified, lay the foundation for the design of \sol. 

Consider two neural networks, $F_{\theta}^1$ and $F_{\theta}^2$, where they are trained on the data distribution $\mathcal{D}$. $F_{\theta}^1$ is a larger primary model, and $F_{\theta}^2$ is a significantly smaller model trained on the same primary task.{\it We speculate that the posterior vector of the smaller model ($F_{\theta}^2$) can be used to detect the impact of a malicious actor on the bigger model ($F_{\theta}^1$)}, even if they disagree on the prediction. This is because the decision boundarys of the $F_{\theta}^2$ significantly overlaps with the decision boundary of $F_{\theta}^1$. An adversary will either modify the decision boundary or the sample distance from the decision boundary in $F_{\theta}^1$ to conduct the attack. Hence, the posterior vector of $F_{\theta}^2$ can be used to detect these modifications.
To test our speculation we orchestrate prediction-switching attacks, \ie adversarial attacks in which malicious actors modify the posterior vectors of $F_{\theta}^1$ during inference. We observe the divergence between the posterior vector of $F_{\theta}^1$ and $F_{\theta}^2$ and assess if we can establish some trends that can be utilized to detect the presence of a malicious actor.
%
As such, we analyze the divergence density distribution of $F_{\theta}^1$ and $F_{\theta}^2$--both before and after conducting a prediction-switching attack against $F_{\theta}^1$. In comparing the divergence trends, \ie $D(F_{\theta}^1(x), F_{\theta}^2(x))$, we consider the following cases:

\begin{itemize}
    \item Natural Agreement {\bf (Case A)}: happens if $argmax(F_{\theta}^1(x)) = argmax(F_{\theta}^2(x))$ prior to any adversarial modifications to the prediction.
    \item Natural Disagreement {\bf(Case B) }: happens if  $argmax(F_{\theta}^1(x)) \neq argmax(F_{\theta}^2(x))$ prior to any adversarial modifications to the prediction.
\end{itemize}
Using these two cases, we aim to recognize identifiable patterns between naturally occurring disagreements and those resulting from malicious actions. In our analysis, we employ three datasets: CIFAR-10, CIFAR-100, and ImageNet, and three architectures: ResNet, DenseNet, and EfficientNet (more details in Section~\ref{sec:evalution}). For assessing divergence, we use Jeffreys' Divergence and Wasserstein-1 measures.
Figure~\ref{fig:JSD} shows the JD distributions between the output of the two models across all datasets and architectures. We identified evident trends between pre-attack and post-attack JD distributions:
\begin{figure}[!t]
\centering
\subfigure{
    \includegraphics[width=\columnwidth]{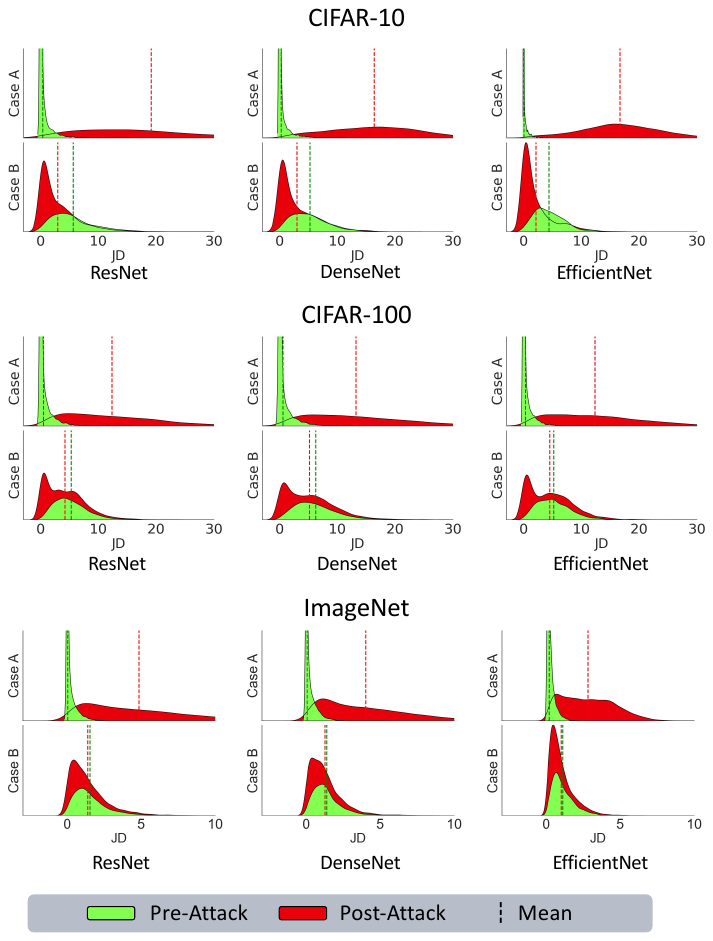}
    \label{fig:cifar10_resnet}}
%
%
%
\vspace{-0.3in}
\caption{The JD measurements of two models. For Case A, the attack increases the divergence value, whereas in Case B (disagreement), the attack decreases the divergence value. Thus validating our reasoning for divergence's role in attack detection.}
\label{fig:JSD}
\vspace{-0.3in}
\end{figure}

\noindent
{\bf Trend 1:} Consider Case A, where both models naturally agree on the predictions before the attack. When the two models agree with each other, and neither of them is under attack, the distribution exhibits small standard deviation and mean values, as depicted by the green curve in Case A. In contrast, when one of the models is subjected to an attack, the JD distribution undergoes a significant transformation, with a sharp increase in both the mean and standard deviation values (red curve in Case A).
%

\noindent
{\bf Trend 2:}Consider Case B, where the two models predict different classes before the attack. We notice a contrasting transformation once one of the models is attacked.
Specifically, when the system is not under attack, the differences between the two models' outputs result in a wide JD distribution with a relatively larger standard deviation and mean values, as shown by the green curves in Case B. In contrast, the attack leads to a narrow distribution with smaller standard deviations and mean values (red curves in Case B).
%

%
%

The trends we have identified, which remained consistent across various architectures and datasets, and the observations from Figure~\ref{fig:JSD} are the key supporting elements that empirically validate our intuition behind the impact of an attack on the similarity of two probability distributions. In particular, the results confirm that \textit{if the two models agree with each other, an attack will lead to an increase in the mean and standard deviation of the divergence, resulting in a wide distribution. In contrast, when the two models have natural disagreements, the attack decreases the mean and standard deviation of the divergence distribution, leading to a narrow distribution.} 
%
%
We also measured the similarity of the two models using the Wasserstein-1 metric to show independence from any particular similarity measure. Our analysis shows similar trends in different value ranges (refer to Table~\ref{tab:JSD_WM_A} and Table~\ref{tab:JSD_WM_B} in Appendix~\ref{apn1}). The observations shown above, however, pose a major challenge. There are overlaps between the pre-attack and post-attack distributions across all cases. Moreover, increasing the complexity of the dataset and the dimensionality of the output vector leads to an increase in the overlap. This growing overlap and complexity serve as indicators of a more complex decision boundary for the attack discrimination task, which can potentially be realized through a neural network. We will address this challenge in our design (Section~\ref{sec:detection}).

\section{Background}
\label{sec:background}
%
In this section, we introduce the concepts relevant to our work, including a trusted execution environment, ML compression techniques, and divergence-based similarity measures.
\subsection{Trusted Execution Environments}
A Trusted Execution Environment (TEE) is an isolated processing environment in which each application is allocated a region of memory, called an enclave, which is protected even from processes running at higher privilege levels~\cite{asokan2014mobile,SabAchBou15}. 
The TEE guarantees the integrity of the run-time states, the confidentiality of the code and data stored on persistent memory, and the authenticity of the executed code and its correct execution via remote attestation between the data owner and the secure enclave. 
%
%
%
The two common TEE implementations are Intel SGX~\cite{costan2016intel} and ARM TrustZone~\cite{alves2004trustzone}. 
\subsection{Model Compression}

\noindent
{\bf Model Quantization}~\cite{han2015deep} is a strategy for model compression and inference process acceleration. Quantization reduces the precision of the model's weight, gradient, or activation values by converting floating point numbers to lower precision integer numbers~\cite{zhou2016dorefa}. Thus, reducing the storage needed for the model and speeding up the inference task~\cite{zhu2016trained}. 
The two primary quantization strategies include post-training quantization, where the weights of a trained model will be compressed (\eg 32-bit float to 8-bit integer), and quantization-aware training, in which the model's weight and gradients are quantized during the training process~\cite{CheWanZho17}.
%
%
%

\noindent
{\bf Knowledge distillation} (KD)~\cite{HinVinDea15} is a machine learning technique for transferring knowledge from a complex neural network(s) (\ie teacher model(s)) to a single model (\ie student model), aimed at model compression, robustness, and performance enhancement~\cite{FurLipTsc18,BagHorRas18,HofAltSch20}. The distillation process entails training the student model using the output feature maps of the teacher model, \ie soft labels, and their corresponding true labels.
%
%
The distilled models are more robust against adversarial attacks when compared with their teacher models~\cite{OrtModMoo21,MarOrtFro22}. Defensive distillation~\cite{papernot2016distillation} is a technique that helps in reducing the effectiveness of the adversarial example attacks against DNNs. 
%
%

\subsection{Similarity Measures}
\label{sec:measures}
Divergence-based similarity measures are statistical techniques to assess the similarity of a probability distribution ($P$) with respect to the reference probability distribution ($Q$). We review the Jeffreys' divergence (JD) and Wasserstein distance.

\noindent
{\bf Jeffreys' Divergence.}
The JD is the bounded symmetrization of the Kullback–Leibler divergence (KLD), which is used to quantify the similarity between two probability distributions $P$ and $Q$.
Unlike KLD, the JD distance represents a normalized score that is symmetric, \ie $D_{J}(P \parallel Q) = D_{J}(Q \parallel P)$ and can be calculated as:
\vspace{-0mm}
$$D_{J}(P \parallel Q) = \frac{1}{2}D_{KL} \Big(P \parallel Q \Big) + \frac{1}{2}D_{KL} \Big(Q \parallel P \Big),$$
%
where $D_{KL}(P \parallel Q)$ is the KLD between $P$ and $Q$.
JD provides a smoothed and normalized score compared to KLD.
%
%

\noindent
{\bf Wasserstein-1 Metric.}
Wasserstein-1 distance is a measure of the distance between two probability distributions. Informally, the Wasserstein distance can be interpreted as the minimum energy cost of moving the mass from x to y to transform probability distribution $P$ to $Q$. Wasserstein-1 metric ($D_W^{1}$) between cumulative distribution functions $P$ and $Q$ is defined as the infimum, taken over the set of all joint distribution $\Gamma(P,Q)$. Thus, the Wasserstein-1 distance between $P$ and $Q$ ($D_W^{1}(P,Q)$) is:
\vspace{-1mm}
$$D_W^{1}(P,Q) =  \inf_{\gamma\in\Gamma(P,Q)}\mathbb{E}_{(x,y)\sim\gamma}[||x-y||].$$

%
%
%

\section{Models and Assumptions}
\label{sec:models}
%
In this section, we first describe our system model and then formalize our threat model and discuss security assumptions. Our system comprises edge servers, ML service providers, and clients. Service providers offer a range of resource-intensive machine learning services, such as image annotation or video analytics, which require input data from the clients. Given the complex nature of the computation load and the need for real-time computation of these services (\eg autonomous driving, multi-player gaming, and traffic monitoring), clients use the services running on the edge server rather than the distant Cloud. We envision a subset of the servers to be equipped with low-end trusted hardware like Intel SGX or AMD TrustZone. 
%

%
%

\subsection{System Model}
\label{system}
Our system comprises a computing ecosystem, service providers, and clients. We consider the quickly growing pervasive edge computing (PEC) environment~\cite{TouSriMis20} as our computing ecosystem. The PEC ecosystem emphasizes the inclusion of client-owned devices in the pool of computing resources, allowing them to carry out clients' requests. Thus, resulting in extensive heterogeneous resource pools in the proximity of clients. 
In the PEC ecosystem, a server is either a pre-deployed infrastructure server or a client's device, such as smartphones and laptops, which runs outsourced services, aiming to generate revenue. In contrast to the infrastructure servers that are static in nature, the client's resources can intermittently join and leave the pool of computing resources.
%
In our system, service providers offer a range of services to clients. We consider resource-intensive machine learning services like image annotation or video analytics, which require input data from the clients. Due to the need for quick computation, the client utilizes edge servers instead of using a distant Cloud.
%

We also consider an access control service such as the recently proposed APECS~\cite{DouTouPan21} for the servers and clients to mutually authenticate each other before outsourcing ML task. This construction prevents unauthorized users from accessing the services and also prevents unauthorized servers from accessing the clients' data. 
%
%
%
Considering the heterogeneity of the PEC ecosystem, we envision a subset of the servers to be equipped with low-end trusted hardware like Intel SGX or AMD TrustZone. 
%


\subsection{Threat Model}
\label{MLTerm}

\label{threat}
%
Attacks against computation outsourcing may target confidentiality, integrity, availability, and privacy. Among all, we focus on threats to integrity.
Thus, the primary objective of \sol is to provide correctness verifiability for the output of the MLaaS inference task. More specifically, given an ML task and its input data, the goal is to enable the client to assess the reliability of the result and infer, with high probability, the trustworthiness of the executed service by the edge server. 

We formalize the threat model by introducing a \textit{security game}, in which the adversary $\mathcal{A}$ generates malicious predictions and deceives the challenger $\mathcal{C}$ into believing that the generated predictions belong to the universe $\mathcal{U}$ of benign input, prediction pairs. 

{\bf Detection Game:} The game utilizes a data universe $\mathcal{U} \coloneqq \{\big (x_{i}, F_{\theta}(x_{i})\big )\}^N_{i=1}$, where  $F_{\theta}:x_{i}\rightarrow[0,1]^n$. The function $F_{\theta}$ is realized through a machine learning model and trained using algorithm $\tau_{F}$. While both the challenger $\mathcal{C}$ and Adversary $\mathcal{A}$ have complete knowledge of input distribution $\mathbb{D}$, trained model $F_{\theta}$, training algorithm $\tau_{F}$, and attack detection algorithm $\tau_{F}^D$, the challenger has no knowledge of the attack algorithm $\tau_{F}^A$ used by adversary. 
\begin{enumerate}
    \item The challenger samples the dataset $\mathcal{D} \subseteq \mathbb{D}$ and trains model $F_{\theta}\leftarrow\tau_{F}(D)$, where $F_{\theta}:x^N_{i=1}\rightarrow y^N_{i=1} \; and \;{(x_i,y_i)}^N_{i=1} \in\mathcal{U}$.
    \item The challenger gives adversary a white-box access to $F_{\theta}$.
    \item The adversary runs $\tau_{F}^A:(x,y)\rightarrow\tilde{y}$, such that $(x,\tilde{y})\cap \mathcal{U}=\phi$, and returns \{$(x,y),(x,\tilde{y})$\}.
    \item The challenger generates a guess $z\in$\{$y,\tilde{y}$\} by running:\\ $\tau_{F}^D(\{y,\tilde{y}\};x,F_{\theta})$.
    \item The adversary wins if $(x,z)\not\in\mathcal{U}$.

\end{enumerate}

The adversary can implement $\tau_{F}^A$ in several different ways, including orchestrating Trojan attacks on image or text classification data~\cite{zhang2021trojaning,cheng2021deep}, performing attacks on the input image in the form of adversarial perturbation~\cite{papernot2016transferability,tao2018attacks}, or leveraging their access to the model and directly modifying the model weights or the prediction vector. 

We do not restrict the adversary's access to the \tea model, which is running in the insecure region of the general purpose processor. The adversary has complete knowledge of the architecture and parameters of the \tea model. We assume the adversary knows the deployed defense mechanism or its components within a trusted enclave but does not have access to any computations taking place inside the enclave. So, they can use the knowledge of the defense algorithm $\tau_{F}^D$ to orchestrate targeted attacks.
\subsection{Attack Modeling}
\label{sec:attack}
%
The primary outcome of the attack we consider in this work is the integrity violation of the MLaaS inference task. To achieve this goal, the adversary can use different techniques. In what follows, we discuss three prominent attack orchestration techniques and implement them in our evaluation. 
All the attacks take place at the edge server, with the attacker deciding which of the different attacks to execute. The attacker's goal is to have the \tea model classify an input with a wrong label. We note that while the adversary can completely control the execution of the \tea model, the adversary cannot interrupt the \stu model as it runs in the secure enclave upon the client’s request.
We implemented all these attacks and tested \sol against them (refer to Section~\ref{sec:evalution}).
%
%

\noindent 
{\bf Prediction Switching Attacks.} In the following attack methodology, the adversary only targets the posterior vector of the deployed machine learning model and modifies it directly, aiming to cause an incorrect prediction. We call this attack {\it naive prediction switching attack} if the attacker modifies or generates the predictions arbitrarily, without any attempt to avoid the detection mechanism. With the knowledge of a detection mechanism, the adversary may deploy different strategies, such as averaging the two highest probability values $p_1$ and $p_2$ ($\mu = \frac{p_1 + p_2}{2}$) to switch the prediction by assigning $\mu + \epsilon$ to the second class (\ie wrong prediction) and $\mu - \epsilon$ to the first class (\ie true prediction), using a small $\epsilon$ value, \eg $0.01\%$ of $\mu$. Alternatively, the adversary trains its own \stu model and uses it to minimize the distance between the forged incorrect prediction and the actual prediction of the service model. We call these {\it advanced prediction switching attacks}.

\noindent
{\bf Well-known Attacks.} 
The adversary can use popular backdoor or adversarial sampling techniques to modify the predictions. For our evaluation, we consider the following attacks:

{\it Fast Gradient Sign Method (FGSM)}~\cite{GooShlSze} perturbations are crafted by calculating the loss between the prediction and the true label. Using the calculated loss, FGSM creates a max-norm constrained ($\epsilon$) perturbation. Given image $x$, the adversarial image $x^{adv}$ can be calculated as $x^{adv} = x + \epsilon \times \textsc{sign}\big(\nabla_x J(\theta,x,y_{true})\big)$.

{\it Projected Gradient Descent (PGD)}~\cite{KurGooBen16} acts as iterative extension of FGSM. The adversarial image is crafted by repeatedly adding perturbation, guided using the loss between the prediction and the target class. Each step of adversarial image generation can be formulated as $x_{N+1}^{adv} = Clip_{X, \epsilon}\big(x_{N}^{adv} + \textsc{sign}(\nabla_x J(\theta,x,y_{target}))\big),$ where $\; x_{0}^{adv} = x $.

{\it Trojan Attacks:} This attack is conducted using the \tea model in a white-box setting. In the context of our work, the adversary can train the \tea model with a poisoned dataset containing trigger embedded images~\cite{CheLiuLi}~\cite{GuDolGar17}. Alternatively, the adversary adds more layers to the \tea model as a Trojan module and trains it using the poisoned data~\cite{TanDuLiu20}, which gets triggered when the input image contains the embedded trigger. 
%
%

%
\begin{figure*}[!t]
\centering
\includegraphics[width=\textwidth]{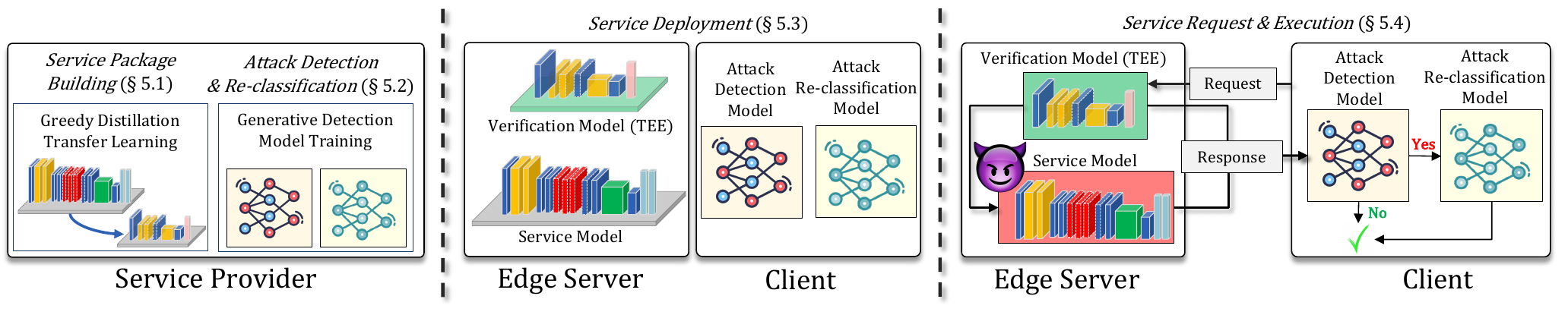}
\vspace{-0.3in}
\caption{In \sol, the provider builds the service package (\S~\ref{sec:prepare}) and the attack detection and re-classification pipeline (\S~\ref{sec:detection}) for deployment on the server (\S~\ref{sec:deploy}) and the client, respectively. The client then sends the service request to the edge server and verifies the result using the attack detection pipeline (\S~\ref{sec:offload}).}
\label{Fig_overview}
\vspace{-0.1in}
\end{figure*}
%
%
%
%
\section{Design of \sol}
\label{sec:design}
%
In a nutshell (refer to Figure~\ref{Fig_overview}), the process of \sol starts with a service provider preparing the service package (\S~\ref{sec:prepare}), including the ML application, \ie \tea model, and its verification component, \ie \stu model, for deployment on the edge servers (the server-side component).
The service provider also builds an attack detection and re-classification pipeline, that includes two neural networks (\S~\ref{sec:detection}), one for attack detection and a second for re-classification. In training the attack detection and re-classification models, the provider uses a generative adversarial network approach.
The service deployment process involves loading the \tea and \stu models to a server with which a trust relationship has already been established (\S~\ref{sec:deploy}). 
After deployment, the server accepts requests for verification from a client. On receiving a client's request, the server runs the \stu model inside the secure enclave and the \tea model in the unprotected region of the processor and returns both outputs to the client (\S~\ref{sec:offload}). 
Finally, the client runs \sol' detection and re-classification functionalities (the client-side component) to identify a potential attack and rectify the outcome of the MLaaS inference as needed. Without loss of generality, we use Intel SGX as the TEE in our implementation reference and explain details using SGX terminology. 
%
\subsection{Service Package Building}
\label{sec:prepare}
\begin{figure}[t]
    \vspace{-0.15in}
    \centering
    \includegraphics[width=0.9\columnwidth]{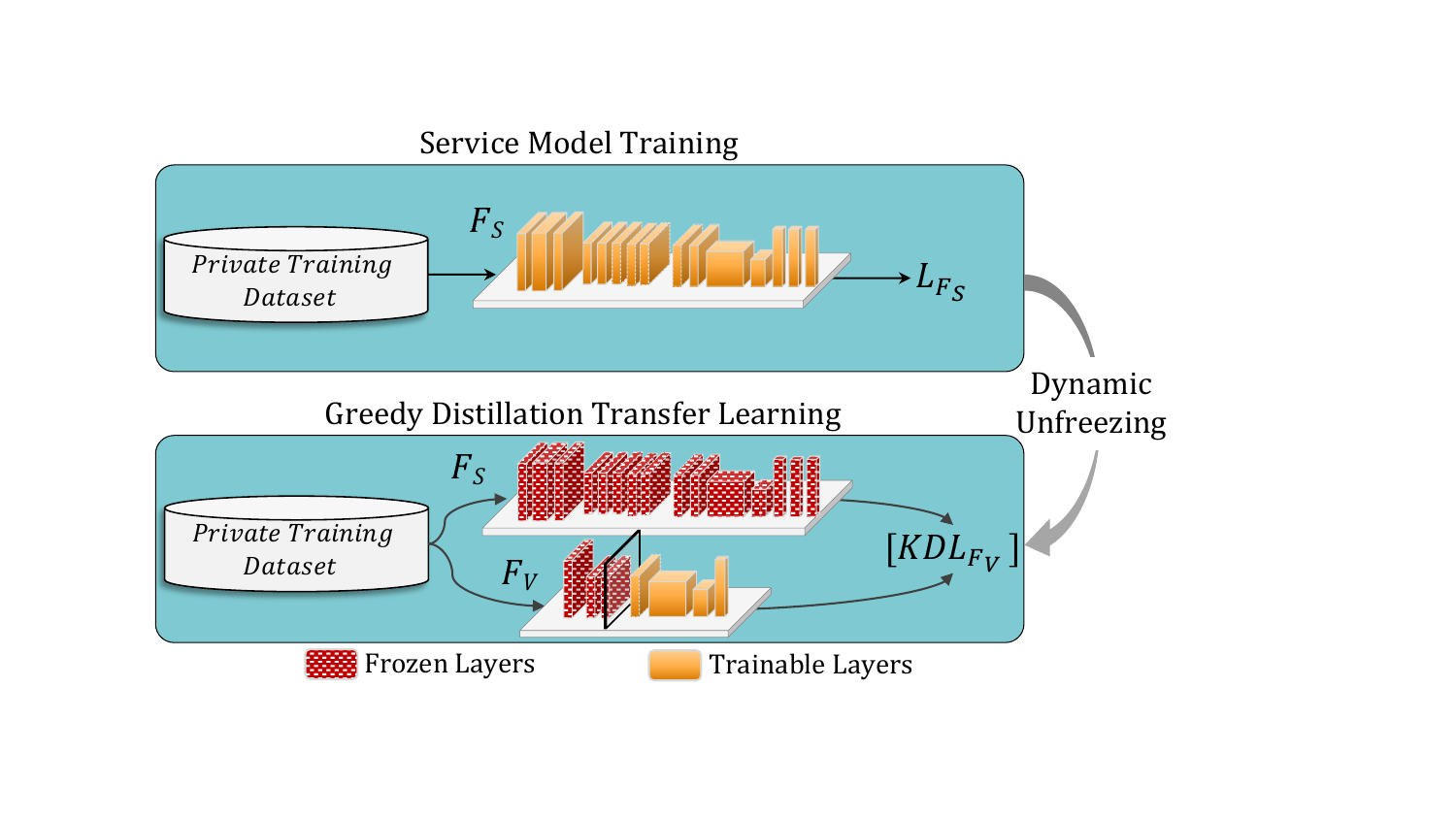}
    \vspace{-0.12in}
    \caption{In service package building, the provider trains the \tea model and uses it to run the greedy distillation transfer learning process, which builds the customized compressed \stu model.}
    \label{Fig_servicePreparation}
    \vspace{-0.3in}
\end{figure}

A service package includes two components--the application (\ie \tea model) and a verification tool (\ie \stu model), which is a small ML model for the verification of its corresponding \tea model. As shown in Figure~\ref{Fig_servicePreparation}, service package development is a two-step process in which the provider first trains its \tea model, \eg image classification, and then dynamically re-train layers of the \stu model using the fully trained \tea model.

As per Algorithm~\ref{alg:distillation}, the service building process takes an untrained \tea model (${F}_S$), an independently pre-trained \stu model (${F}_V$), the privately owned training set ($\mathbb{D}_T^{Priv}$), and a set of hyper-parameters ($\lambda$ and $\alpha$). Upon completion, it returns the fully trained \tea (${F}_S$) and \stu (${F}_V$) models. Training the \tea model follows the standard training process using $\mathbb{D}_T^{Priv}$ with the defined loss function for as many epochs as required (Lines~1-3). 
%
The \stu model is a distilled and smaller version of the \tea model, so it renders minimal computation overhead when deployed into the edge server's enclave. 
To reduce the cost of knowledge distillation, we propose {\it Greedy Distillation Transfer Learning} (\ddf), a distillation technique that results in a time-efficient procedure for training the small \stu model.

The \ddf process takes a pre-trained model that is significantly smaller as compared to the \tea model and adaptively unfreezes the layers that require re-training and fine-tuning. More specifically, \ddf first splits the \stu model starting from its last layer (Line~8). Given the model split, \ddf runs ${F}_S$ and ${F}_V$ on $\mathbb{D}_T^{Priv}$ to obtain the soft labels of the \tea model and the partially trained \stu model (Lines~8-10). In Line~10, \ddf calculates the knowledge distillation loss (\ie KDL) of these values and the one-hot encoding ($\tau$) of the labels (Line~6). Note that the weight of the average ($\alpha$) changes over time. \ddf starts with a larger $\alpha$ value, giving more weight to the teacher model, and then adaptively decreases it over time, giving higher weight to \tea true label. 
The rationale for adaptively changing $\alpha$ is to initially guide the model towards the \tea model and then decrease it for it to improve over samples the teacher model is providing incorrect prediction.
\ddf then uses the loss value and the stochastic gradient descent method to update the trainable layers (Line~11) using $\mathbb{D}_T^{Priv}$.
%

%

%
%

%
\setlength{\algomargin}{1.5em}
\begin{algorithm2e}[!t]
\caption{Service Package Development}
\label{alg:distillation}
\raggedright
\KwIn{${F}_S$ (untrained), ${F}_V$ (public pre-trained), $\mathbb{D}_T^{Priv}, \lambda \in [0,1]$, $\alpha \in [0,1].$}
\BlankLine
\KwOut{$\langle F_S, F_V \rangle.$}
\BlankLine
\BlankLine
\nonl{\bf Service Model Training:}

\nl\For{number of the training epochs}
{
    Use stochastic gradient descent to update ${F}_S$ on:
    $\mathcal{L}_{{F}_S} \big({F}_S(x_i),y_i \big), \forall (x_i,y_i) \in \mathbb{D}_T^{Priv}$
}
{\bf Store} $F_S$ \Comment{$F_S$ is the fully-trained Service model.}
$\Bigcdot_{j=1}^n {F}_{V}\leftarrow{F}_{V1}\Bigcdot{F}_{V2}\Bigcdot\cdots\Bigcdot{F}_{Vn}$
\BlankLine
\BlankLine
\nonl{\bf Distillation-based Fine-tuning:}

\nl$l \leftarrow n$ \Comment{Set the cut-layer to the last layer of ${F}_{V}$.}

$\tau \leftarrow$ \textsc{One Hot Encoding} of data label $(y_i).$

\While{$\big(l \ge 1 \,\,\, \&\& \,\,\, $\textsc{Acc}$({F}_{S}) \ge \lambda * $\textsc{Acc}$({F}_{V})\big)$}
{

        $\{\Bigcdot_{j=1}^{l-1} {F}_{Vn},\Bigcdot_{j=l}^n {F}_{Vn}\} \longleftarrow$ \textsc{ML-Split}$({F}_{V}, l)$
    
        \For{$\big(\forall (x_i,y_i) \in \mathbb{D}_T^{Priv}\big)$}
        {
    
            $KDL\Big(F_{S}(x), {F}_{V}(x), \tau, T \Big) $= 
            $- \alpha \mathbb{E}_{{F}_{S}(x)}\Big[\log{\Big(\frac{{F}_{V}(x)}{T}\Big)}\Big]$
            $- (1-\alpha) \mathbb{E}_{\tau}\Big[\log{\Big({F}_{V}(x)\Big)}\Big]$\\
            
            Use stochastic gradient descent to update $\Bigcdot_{j=l}^n {F}_{V}$ on:
            $KDL\Big(F_{S}(x), {F}_{V}(x), \tau \Big),$ where $gradients = gradients \times T^2 $
        }
        
                    
            Unfreeze $(\Bigcdot_{j=l}^n {F}_{V})$ by adjusting the cut layer in ${F}_{V}$ : $l \leftarrow l-1$.
}
\textsc{Quantize}$(F_V)$ \Comment{Dynamic-range quantization.}

{\bf Return $\langle F_S,F_V \rangle$} \Comment{Fully-trained Service model and fine-tuned distilled verification model.}
\end{algorithm2e}
%

\begin{figure}[!t]
    \centering
\subfigure[Accuracy on CIFAR-10.]{
    \includegraphics[width=0.22\textwidth]{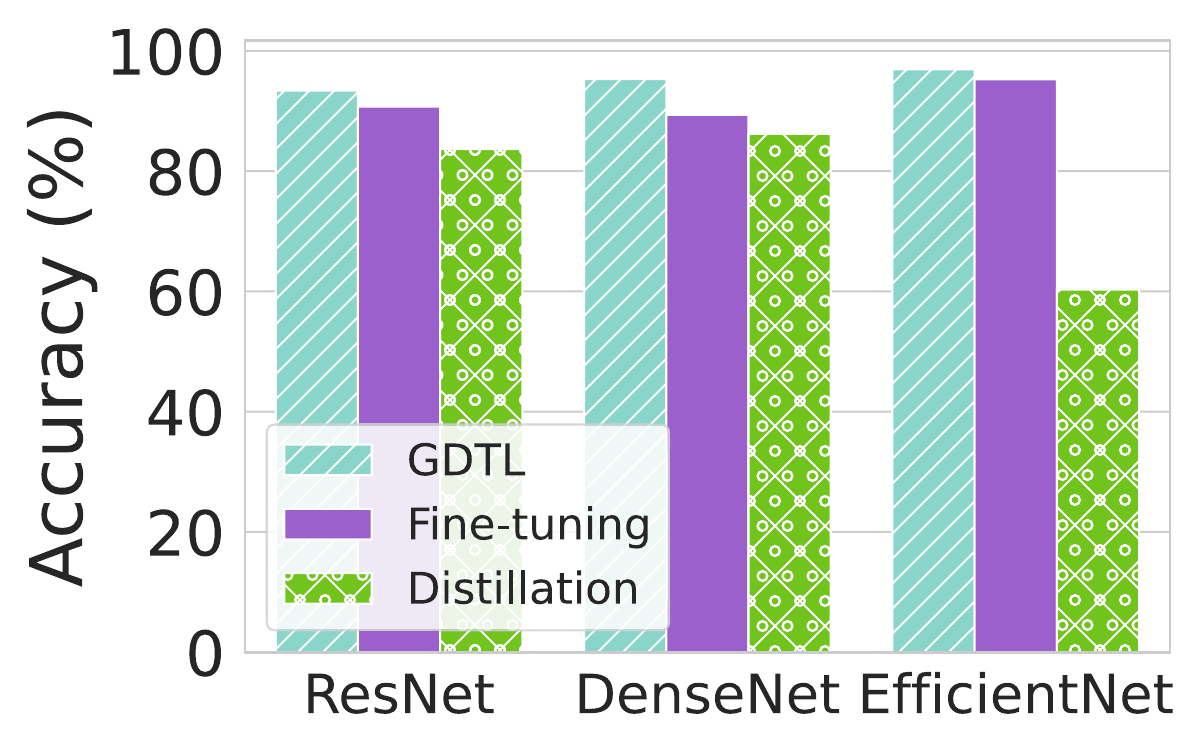}
    \label{fig:distill_acc_10}}
\subfigure[Accuracy on CIFAR-100.]{
    \includegraphics[width=0.22\textwidth]{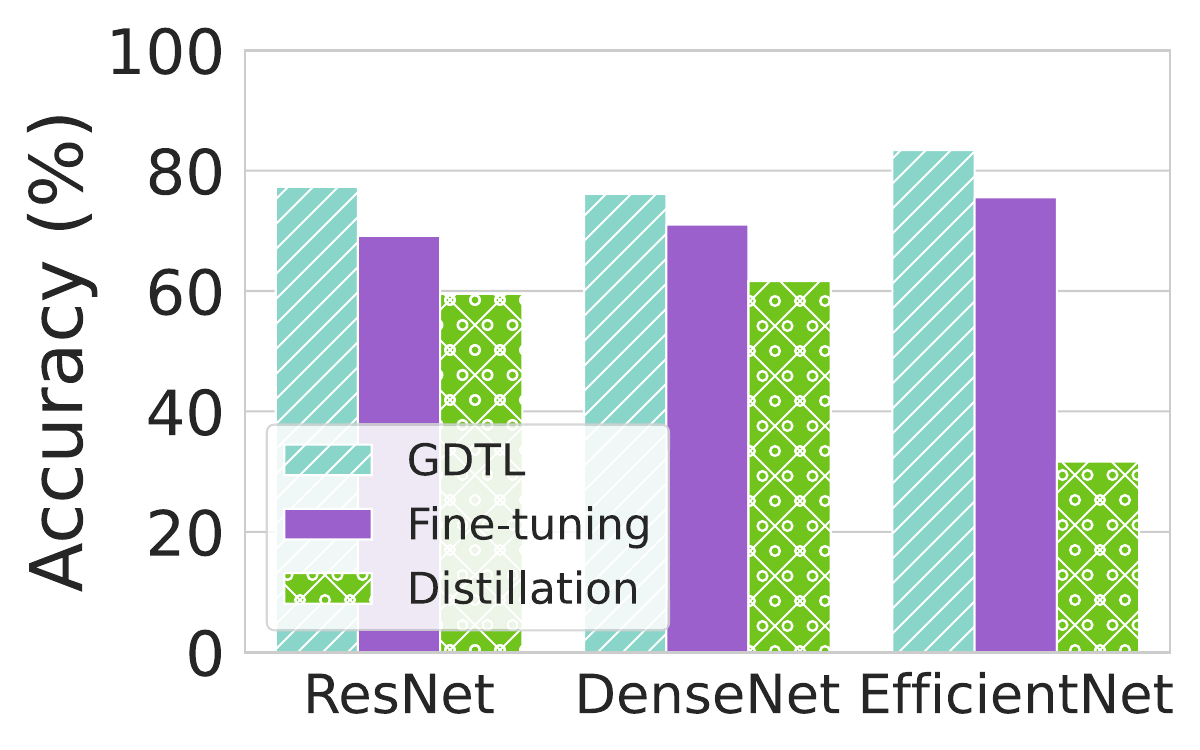}
    \label{fig:distill_acc_100}}
\vspace{-0.1in}
\subfigure[Time on CIFAR-10.]{
    \includegraphics[width=0.22\textwidth]{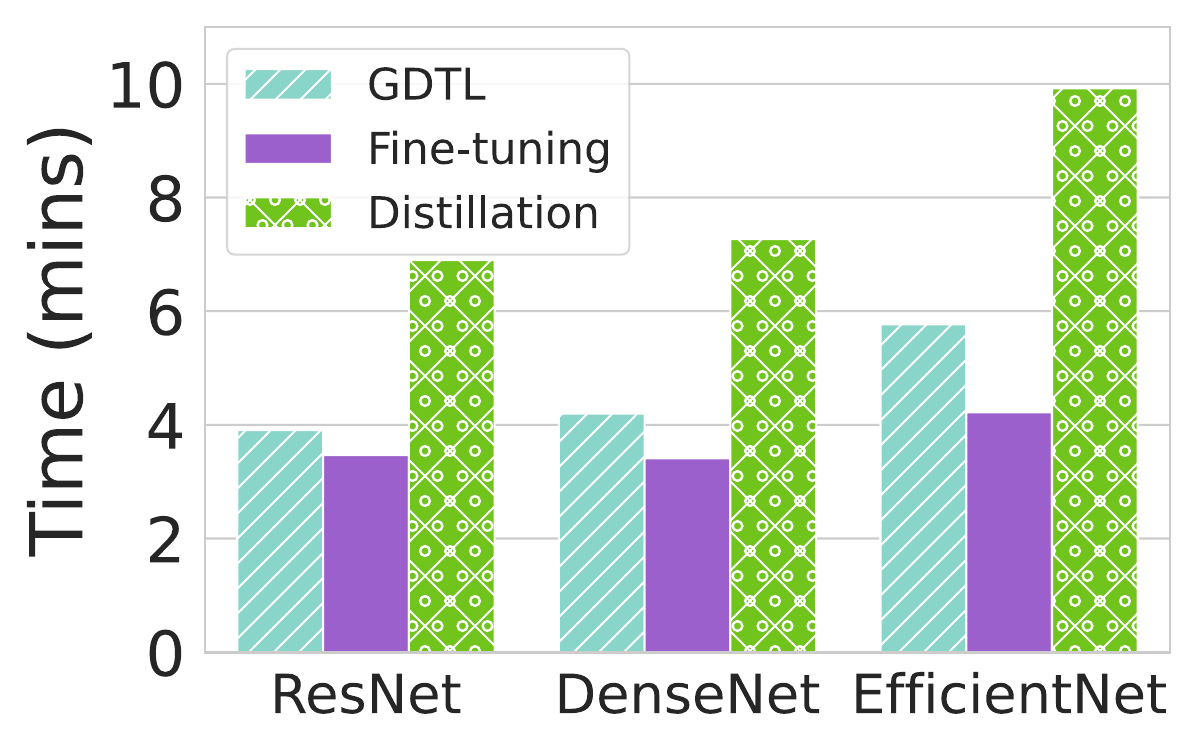}
    \label{fig:distill_time_10}}
\subfigure[Time on CIFAR-100.]{
    \includegraphics[width=0.22\textwidth]{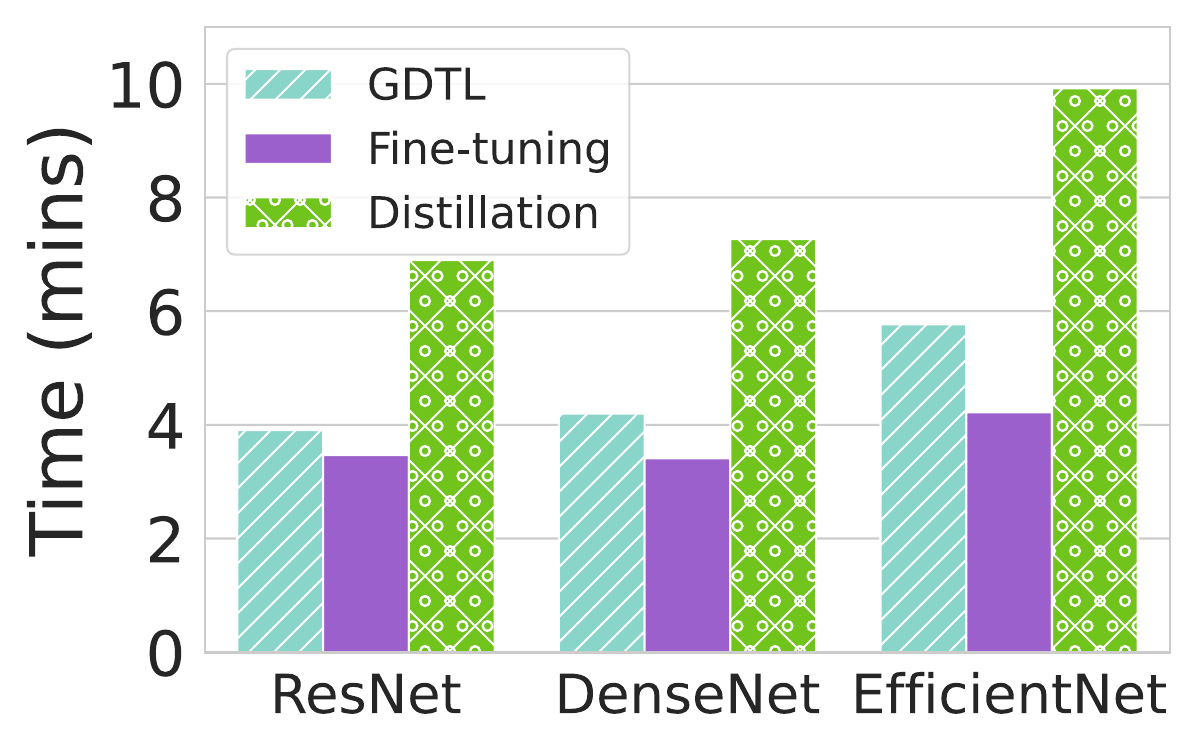}
    \label{fig:distill_time_100}}
\vspace{-0.1in}
\caption{The performance comparison between \ddf, distilling from scratch (Distillation), and fine-tuning (Transfer Learning) across multiple datasets and architectures. The results show the accuracy and per epoch time for all approaches showcasing how GDTL outperforms other approaches while being resource-efficient.}
\label{fig:distillation_perf}
\vspace{-0.15in}
\end{figure}

Each training iteration, \ddf compares the accuracy and precision of the partially fine-tuned \stu model with the \tea model. If the \stu model's accuracy is higher than the $\lambda$ factor of the \tea model's accuracy, the \ddf process stops training ${F}_V$ (Lines~7). Otherwise, \ddf dynamically unfreezes another layer of the \stu model for fine-tuning (Line~12). In essence, the \ddf process gradually splits the model from the last layer towards the first and continuously re-trains and fine-tunes the trainable layers.
We note that $\lambda$ is a hyper-parameter that the provider adjusts based on various factors like its available computing resources and the desired verification accuracy. A $\lambda$ value closer to 1 results in a high-accuracy \stu model but requires more fine-tuning steps. 
Our initial assessment (Figure~\ref{fig:distillation_perf}) shows that \ddf outperforms standard distillation and fine-tuning in terms of accuracy and is comparable to fine-tuning in terms of per-epoch execution time. \ddf also reduces the difficulty of finding suitable regularization, which is the primary reason behind the lower accuracy of classical knowledge distillation~\cite{YanSha20}.

For further compression, \ddf applies model quantization as a common technique for approximating a neural network that uses floating-point numbers using a neural network of lower bit-width numbers. 
Among all, in \sol, we used dynamic range quantization (Line~13), which is a post-training approach that does not need additional model re-training and fine-tuning. Dynamic range quantization converts 32-bit floating point numbers into 8-bit integers, with the resultant model running using floating point operations. 
At the end of this process, the service package is completed. It should be pointed out that the design of the distillation algorithm relies on an underlying assumption that pre-trained weights utilized are trusted. Violation of this assumption can lead to vulnerabilities~\cite{YaoLiZhen19}, which are out of the scope of this work.
%
%
\subsection{Generative Attack Detection and Re-classification Training}
\label{sec:detection}
%

%
Based on the observations made in section~\ref{sec:observation} it is evident that the divergence between the \tea and \stu models' probability distributions reveals statistical information, which is essential for attack detection. Thus, we utilize the differences in the relative divergence as features to assist attack detection.
%
%

For attack detection, a naive approach would be using the resultant divergence of the two probability distributions to set a fixed attack detection threshold. However, manually setting a threshold is not effective since {\it (i)} the threshold is model and data-specific, and {\it (ii)} the attack detection decision boundary is non-linear, for which a linear threshold will be ineffective. 
Supervised learning algorithms can effectively learn such non-linear decision boundaries and improve the detection accuracy. However, these algorithms will utilize known attack signatures and, hence, can become ineffective in detecting variations of known attacks or attacks outside the training dataset. As such, we used a GAN framework for training \sol' attack detection and re-classification models where no predefined attack signatures are used during the training phase, allowing the trained model to be more robust against unseen attack signatures.
%
%

The correlation of the outputs of the \tea and \stu models will result in five possible cases. Case {\em C1} is when both models agree with the ground truth. Case {\em C2} is when only the \tea model agrees with the ground truth while Case {\em C3} represents those instances where only the \stu model agrees with the ground truth. For Case {\em C4} both models make two different incorrect predictions. In Case {\em C5}, both models make the same incorrect prediction, \ie disagree with the ground truth but happen to agree with each other.
These cases are accounted for in our GAN framework to train the attack re-classification model. Note that the re-classification model is a five-class classifier, which returns one of these cases regardless of the number of classes in the \tea model.

\noindent
{\bf GAN-based Attack Detection and Re-classification.}
We formalize the training of ~\sol' attack detection model ($\mathcal{D}$) and the attack generator model ($\mathcal{G}$) as a min-max game. The generator is an attack crafting neural network, which takes samples from the private dataset ($\mathbb{D}_T^{Priv}$) and it returns outputs that are different from the service model. The goal of the detection model, \ie the discriminator, is to differentiate between the output of the service model and the generator model's crafted outputs. The generative training process simultaneously trains a re-classification model ($\mathcal{R}$), which aims at correcting the attack's outcome by reclassifying the output of the service model when under attack.

In training the generator and detection models, we define the objective function of the min-max game as:
\begin{align*}
\begin{split}
\min_{\mathcal{G}}\max_{\mathcal{D}}V(\mathcal{G},\mathcal{D}) & =  \mathbb{E}_{x\sim p_{\text{data}}(x)}\Big[\log{\Big(\mathcal{D}\big(F_S(x), F_V(x)\big)\Big)}\Big] \\ & +  \mathbb{E}_{x\sim p_{\text{data}}(x)}\Big[ \log{\Big(1 - \mathcal{D}\big( \mathcal{G}(x), F_V(x)\big)\Big)}\Big] \\ & -
\mathbb{E}_{x\sim p_{\text{data}}(x)}\Big[\log{\big(1 - \mathcal{G}(x)\big)}\Big].
\end{split}
\end{align*}
The first and second terms are the cross-entropy between the output of the attack detection model and its true label $\mathcal{Y}_\mathcal{D}$ ($\in \{0, 1\}$), where 1 signifies no attack while 0 signifies attack. The third term is the cross-entropy between the generator model's output and the private data's true label.
Our proposed objective function is different from the conventional generative adversarial networks in two aspects. First, the addition of the third term penalizes the generator model every time its prediction is the same as the true label. Thus, preventing the generator model from converging with the service model.
Second, the generator model does not sample any input noise. Instead, it uses samples from the private dataset as input and crafts a malicious output with a wrong label, while mimicking the decision boundary of the service model. 
\begin{figure}[!t]
    \centering
    \includegraphics[width=\columnwidth]{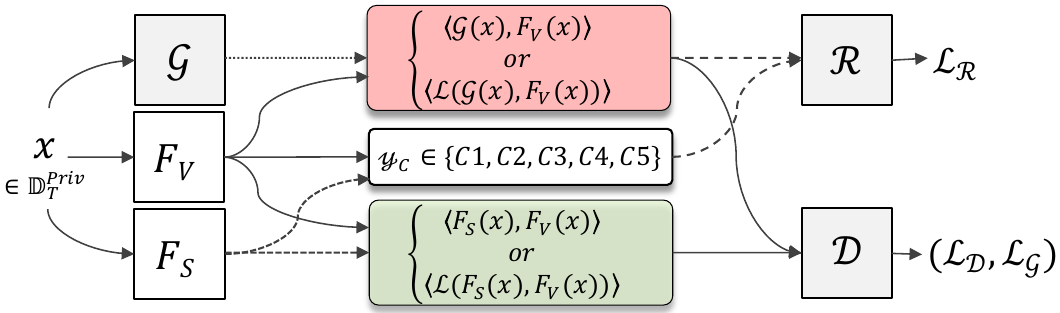}
    \vspace{-0.2in}
    \caption{We formalized the attack detection and subsequent re-classification models as a GAN, in which the attack generator ($\mathcal{G}$) crafts various attacks to train the attack detection ($\mathcal{D}$) model in a two-player min-max game. In the process of training $\mathcal{G}$ and $\mathcal{D}$, the framework simultaneously trains a re-classification model ($\mathcal{R}$), aiming to correct the outcome of the attack.}
    \label{Fig_detection}
    \vspace{-0.1in}
\end{figure}

Recall that the generator model takes the samples from the private dataset as input and generates malicious outputs. We define a malicious output to be a prediction that is different from the true label but has a probability distribution similar to a naturally occurring misclassification. To achieve this behavior, we define the generator model's loss function as:
\begin{align*}
\begin{split}
\mathcal{L}_{\mathcal{G}} = -\log{\Big(\mathcal{D}\big(\mathcal{G}(x), F_V(x)\big)\Big)} - \log{\big(1 - \mathcal{G}(x)\big)}.
\end{split}
\end{align*}
The first term in $\mathcal{L}_{\mathcal{G}}$ rewards the generator model every time the detector model classifies the output of the generator model as a valid output. The second term rewards the generator model for making predictions that are different from the true labels.

The input of the detection model comprises two tuples (Figure~\ref{Fig_detection}). The first tuple consists of the outputs of the generator model $\mathcal{G}(x)$ and verification model $\big(F_V(x)\big)$ alongside their cross-entropy loss $\Big(\mathcal{L}\big(\mathcal{G}(x), F_V(x)\big)\Big)$. The second tuple consists of the outputs of the service model $\big(F_S(x)\big)$ and verification model $\big(F_V(x)\big)$ alongside their cross-entropy loss $\big(\mathcal{L}(F_S(x), F_V(x))\big)$.
We define the loss function of the detection model as the binary cross-entropy between the model's output and its true label:
\begin{align*}
\begin{split}
\mathcal{L}_{\mathcal{D}} &  = -\log {\Big(\mathcal{D}\big(F_S(x), F_V(x)\big)\Big)} \\ & - (1 -  \mathcal{Y}_\mathcal{D}) * \log {\Big(1 - \mathcal{D}\big(\mathcal{G}(x), F_V(x)\big)\Big)}.
\end{split}
\end{align*}
The primary challenge in training the detection model is differentiating the naturally occurring prediction disagreement between the \tea and \stu models and the disagreement caused by the attack. As mentioned earlier the trend we observed in our analysis regarding the difference in the divergence ranges between a natural misclassification and an attack allows the detection model to differentiate between the two (refer to Figure~\ref{fig:JSD} in Section~\ref{sec:observation}).

Per Figure~\ref{Fig_detection}, during the training process, the re-classification model receives the output of the generator model $\big(\mathcal{G}(x)\big)$ and verification model $\big(F_V(x)\big)$ with their cross-entropy loss $\big(\mathcal{L}(\mathcal{G}(x), F_V(x))\big)$. It also takes the compatibility label between the \tea and \stu models, \ie $\mathcal{Y}_c$, which signifies the correctness of the outputs of these models and their similarity.
Using these inputs, we defined the re-classification model's loss as the cross entropy between the output of the re-classification model and $\mathcal{Y}_c$ as:
%
\begin{align*}
\mathcal{L}_{\mathcal{R}} = -\sum_{c=1}^{5} \mathcal{Y}_{c}*\log\Big(\mathcal{R}\big(\mathcal{G}(x), F_V(x)\big)\Big),
\end{align*}
where $\mathcal{Y}_c \in \{C1, C2, C3, C4, C5\}$. 

After training the attack detection and re-classification models using the proposed GAN, the service provider shares them with the clients. Thus, allowing the clients to independently validate the results of the MLaaS inference task, executed by the edge server.
%
%
%
\subsection{Service Deployment}
\label{sec:deploy}
\begin{figure}[!t]
    \centering
    \includegraphics[width=1\columnwidth]{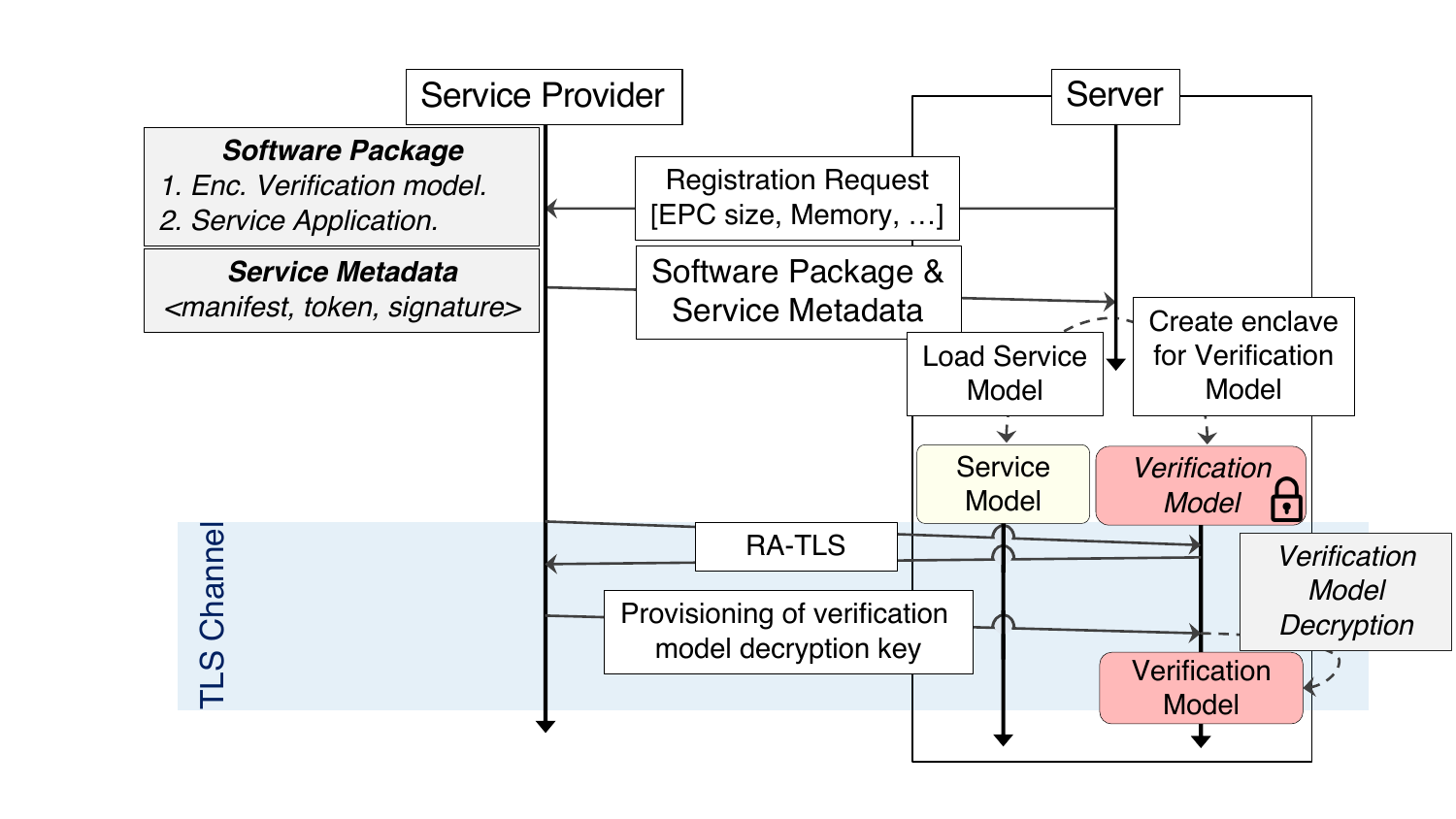}
    \vspace{-0.2in}
    \caption{During the service deployment process, the service provider securely deploys the \stu model on a secure enclave on the remote edge server and provisions the decryption key after remote attestation. Thus, guaranteeing the integrity of the \stu model.}
\label{Fig_deployment}
\vspace{-0.15in}
\end{figure}

Without loss of generality, we consider two possibilities for service deployment. In the first approach, the \pecserver initiates the deployment process by sending a request to the provider with requisite information about its resources, \eg EPC size, memory, bandwidth, and the service(s) it is willing to offer. The choice of the service depends on criteria like service demand or resource availability. Alternatively, the service provider initiates the deployment request to a potential \pecserver based on service demands in the \pecserver's locale. Per Figure~\ref{Fig_deployment}, we adopt the first approach.
%
%

After the final agreement, the provider sends a complete application package to the \pecserver. The application package includes the \tea model, the encrypted \stu model, and the service metadata. The service metadata contains application-specific configuration information, which is necessary for the integrity verification of the ML application. In particular, for Intel SGX, the metadata contains a {\tt .manifest.sgx} file for configuring the secure application environment by the SGX SDK along with a {\tt .token} and a {\tt .sig} file, which SGX uses to verify the integrity of the enclave-loaded application and files. 
The \pecserver then initiates a secure enclave for loading the encrypted \stu model. It then establishes an RA-TLS 
channel~\cite{KnuSteCha18} with the service provider for provisioning the application decryption key into the secure enclave. Thus, allowing the enclave to decrypt the \stu model.

The \servpro shares requisite enclave information, such as $\langle${\tt MR\_ENCLAVE, MR\_SIGNER, ISV\_PROD, ISV\_SVN}$\rangle$ with the clients, enabling them to verify the chain of trust during the remote attestation process (\ie RA-TLS) at the service outsourcing step. The provider generates these values while creating the SGX-supported application and shares them with the client when the application is downloaded. The RA-TLS protocol helps the client trust the output of the \stu model running in the enclave.
\subsection{Service Request and Execution}
\label{sec:offload}
\begin{figure}[!t]
    \centering
    \includegraphics[width=\columnwidth]{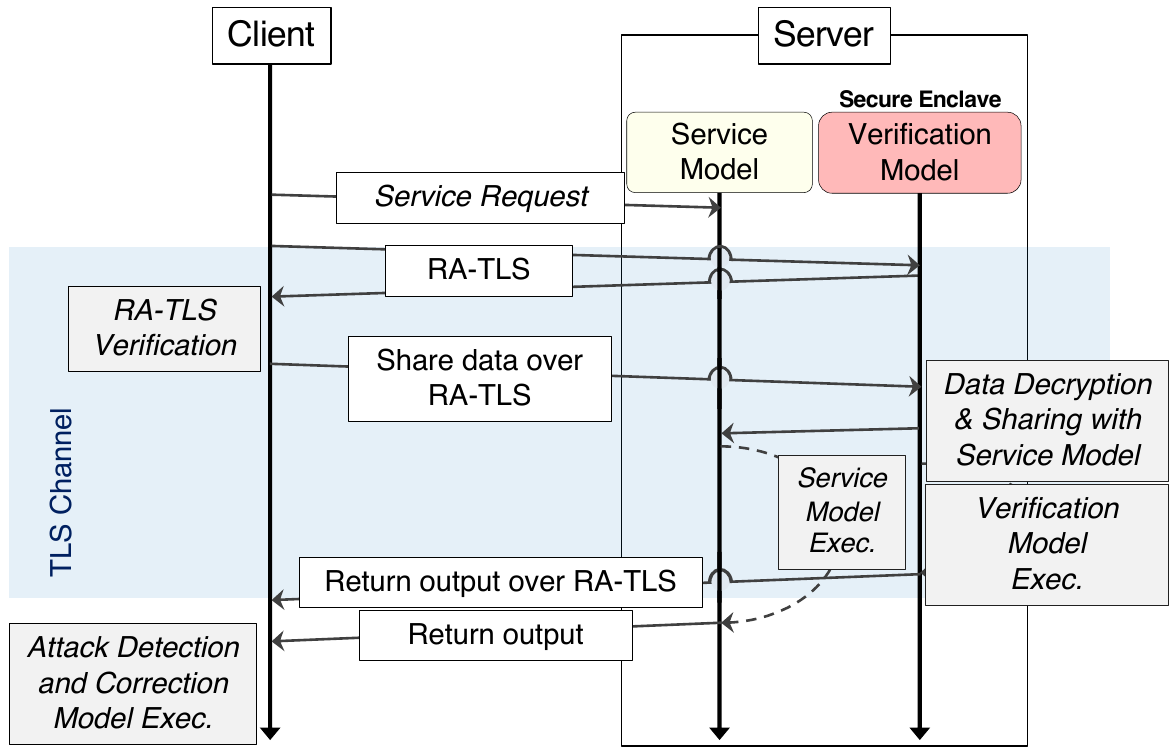}
    \vspace{-0.2in}
    \caption{During service request, the client securely shares the data with the secure enclave running the \stu model, which in turn shares it with the \tea model. Thus, ensuring that the \stu model has access to trustworthy data.}
    \label{Fig_execution}
    \vspace{-0.15in}
\end{figure}

%
For service request, the client starts by discovering the servers that offer the service and have sufficient resources. If the service discovery fails, the client can either request the service from the provider or request a server to deploy the service. We do not discuss service discovery and assume the client finds a nearby server. 
Once the client has identified the server, it sends a service request to the server (Figure~\ref{Fig_execution}). Considering our threat model, a malicious server can change the data received from the client before sending it to the \stu model for verification. Hence, the client first establishes an RA-TLS channel with the enclave to perform remote attestation. The remote attestation process on the client uses $\langle${\tt MR\_ENCLAVE, MR\_SIGNER, ISV\_PROD, ISV\_SVN}$\rangle$ information obtained from the provider to verify the integrity of the respective enclave in terms of its software and configuration. On successful remote attestation, \ie verifying the chain of trust and ensuring that the client is directly communicating with the secure enclave, the client securely shares the input data with the enclave. The application inside the enclave decrypts the client's data, shares it with the \tea model, and executes the \stu model. 
Alternatively, the client can establish a TLS channel with the \tea model and an RA-TLS channel with the \stu model to directly share the data with both. We use the first approach in our experiments.

After completing the ML inference task, the \tea and \stu models return their outputs to the client. Note that the communication between the client and the secure enclave is protected through the established RA-TLS channel. The \tea model can either set up a secure connection to the client and securely share the result or use an insecure channel. The client then uses \sol' detection and re-classification pipeline on the outputs of the \tea and \stu models to verify the correctness of executed ML inference task and detect potential attacks.

\section{Experiments}
\label{sec:evalution}
%
\begin{table}[!t]
    \centering
    \caption{Accuracy of \tea and \stu models.}
        \vspace{-0.1in}
        \begin{tabular}{lcccc}
        \multicolumn{2}{c}{}
                &{\rotatebox[origin=c]{-0}{ResNet}} & {\rotatebox[origin=c]{-0}{DenseNet}} & {\rotatebox[origin=c]{-0}{EfficientNet}} \\ 
        \hline
        \multirow{2}{*}{\rotatebox[origin=c]{0}{CIFAR-10}} 
                & Service & 93.77\% & 94.93\% & 96.74\% \\ 
                \cline{2-5}
                & Verification & 93.38\% & 95.30\% & 96.91\% \\
        \hline
        \multirow{2}{*}{\rotatebox[origin=c]{0}{CIFAR-100}}  
                & Service & 76.99\% & 78.69\% & 83.79\% \\ 
                \cline{2-5}
                & Verification &77.31\%  &76.13\%  &83.41\%  \\
        \hline
        \multirow{2}{*}{\rotatebox[origin=c]{0}{ImageNet}}  
                & Service & 73.01\% & 73.32\% & 79.73\% \\ 
                \cline{2-5}
                & Verification & 70.65\% & 69.74\% & 72.16\% \\
        \hline
    \end{tabular}
\label{tab:model_accuracy}
\vspace{-0.1in}
\end{table}
In this Section, we review our benchmark datasets and models and elaborate on our system and experiment setup. We first analyze the similarity of the \tea and \stu models pre- and post-attack and share our observations, which confirms the rationale behind our design. Finally, we assess the efficacy of \sol in detecting and re-classifying attacks, followed by system performance analysis.
\subsection{Data and Models}
We evaluated \sol\footnote{\sol Code is available on \url{https://github.com/akumar2709/Fides\_AsiaCCS}} using datasets of various complexities:
\begin{itemize}
    \item CIFAR-10 consists of 60K color images ($32\times32$ pixels) with 10 classes and 6K images per class. There are $50000$ training and $10000$ test images~\cite{CIFAR}.
    \item CIFAR-100 features 100 classes with 600 images per class with pixels resolution of $32\times32$. There are 50000 training and 10000 test images~\cite{CIFAR}.
    \item ImageNet-1K contains about 1.3M annotated images (about 1.28M training, 50K validation, and 100K test images), grouped into 1000 classes~\cite{deng2009imagenet}. The resolution of images is $469\times387$ pixels; we cropped these images to $224\times224$ pixels.
\end{itemize}
\begin{figure*}[!ht]
\centering
\subfigure[Detection accuracy on CIFAR-10.]{
    \includegraphics[width=0.3\textwidth]{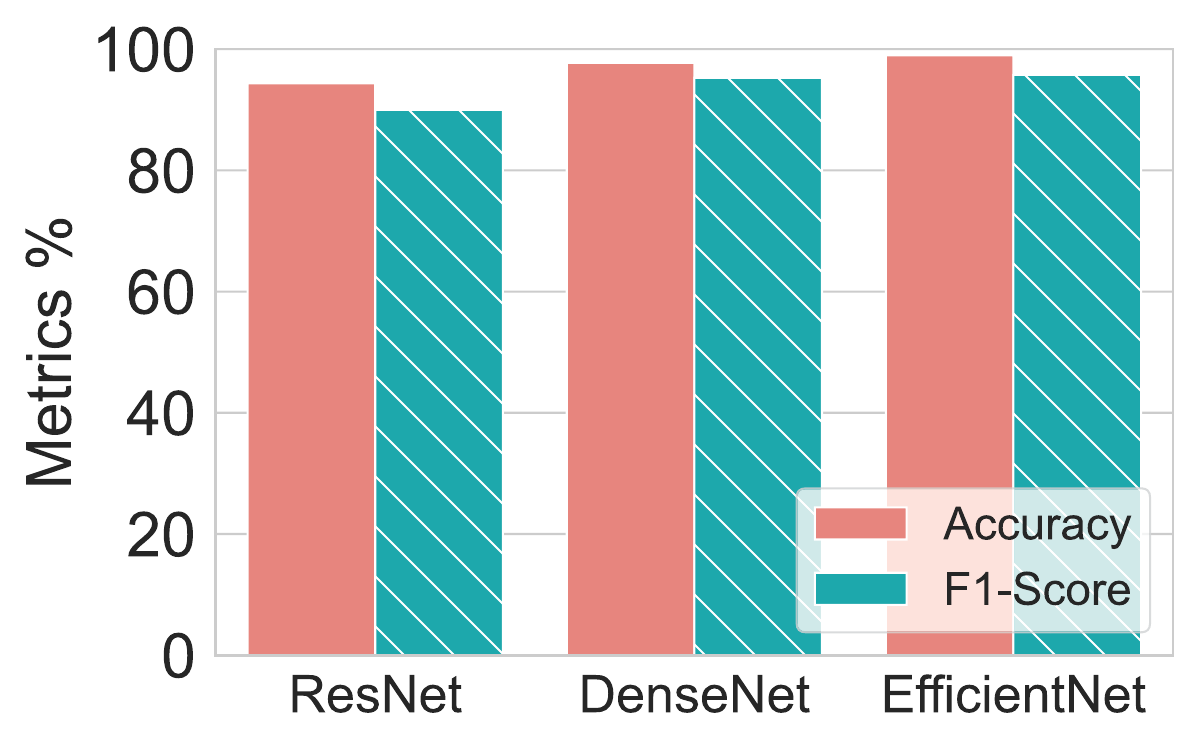}
    \label{fig:detection_cifar10}}
\subfigure[Detection accuracy on CIFAR-100.]{
    \includegraphics[width=0.3\textwidth]{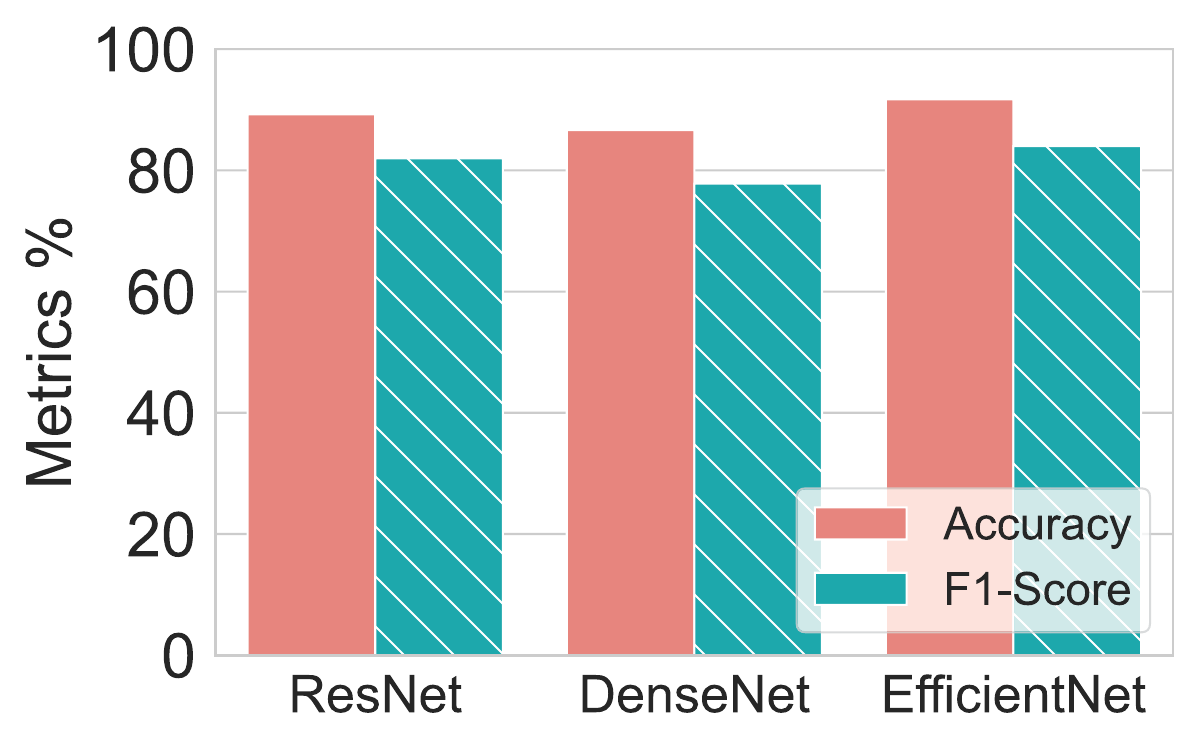}
    \label{fig:detection_cifar100}}
\subfigure[Detection accuracy on ImageNet.]{
    \includegraphics[width=0.3\textwidth]{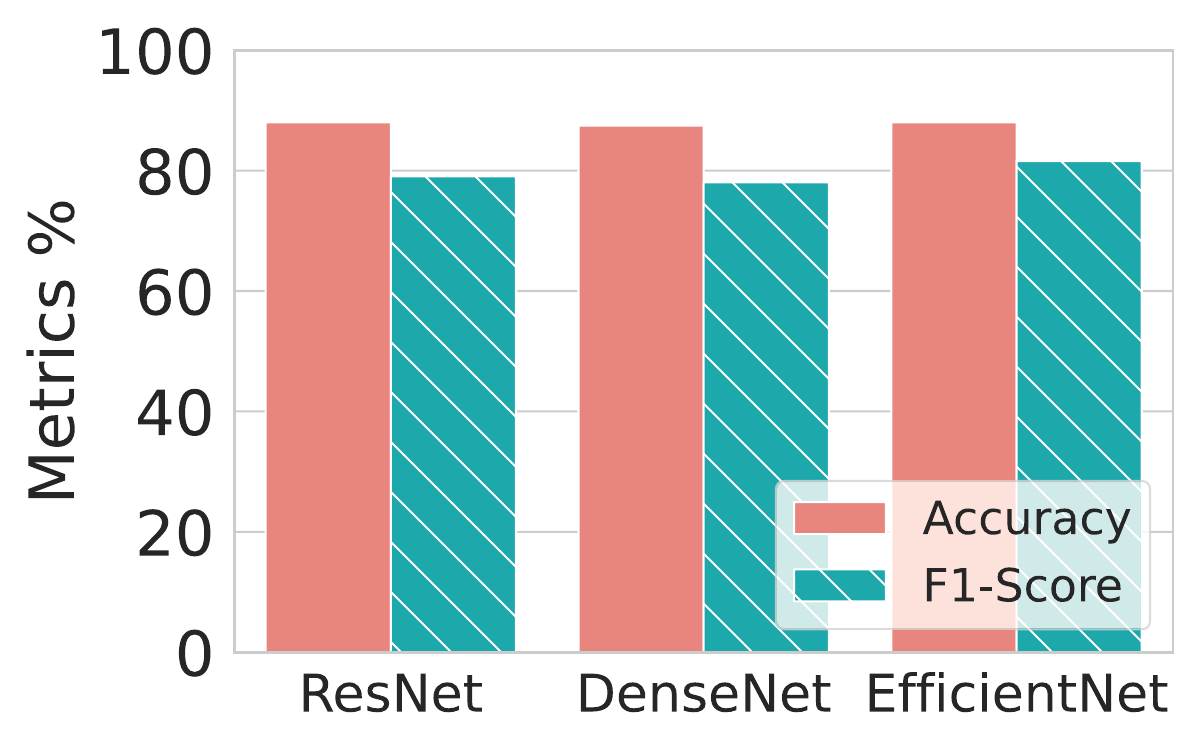}
    \label{fig:detection_imagenet}}
\subfigure[Re-classification accuracy on CIFAR-10.]{
    \includegraphics[width=0.3\textwidth]{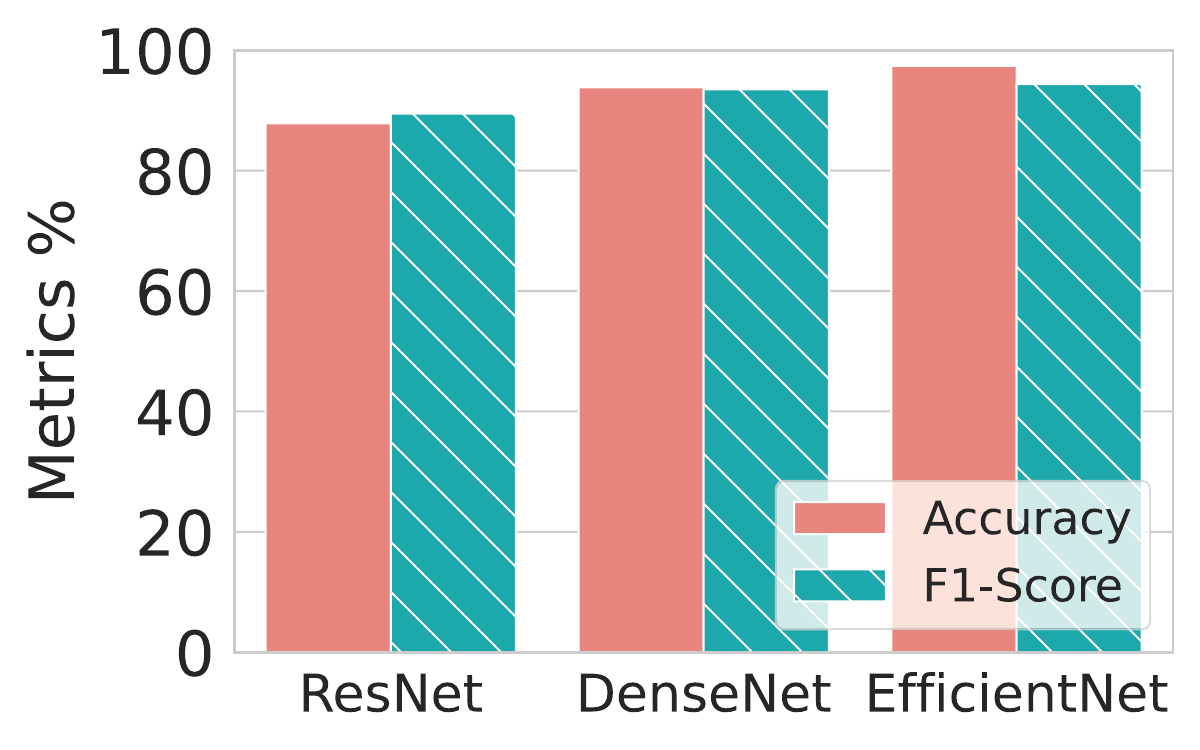}
    \label{fig:correction_cifar10}}
\subfigure[Re-classification accuracy on CIFAR-100.]{
    \includegraphics[width=0.3\textwidth]{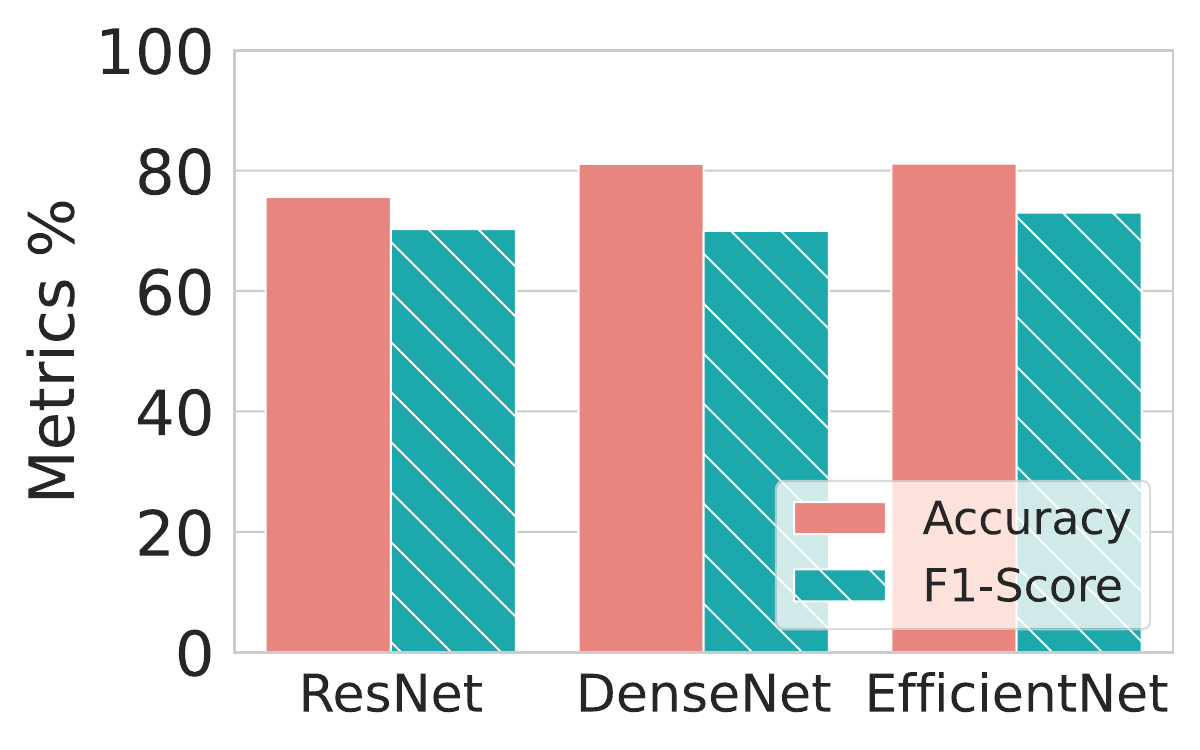}
    \label{fig:correction_cifar100}}
\subfigure[Re-classification accuracy on ImageNet.]{
    \includegraphics[width=0.3\textwidth]{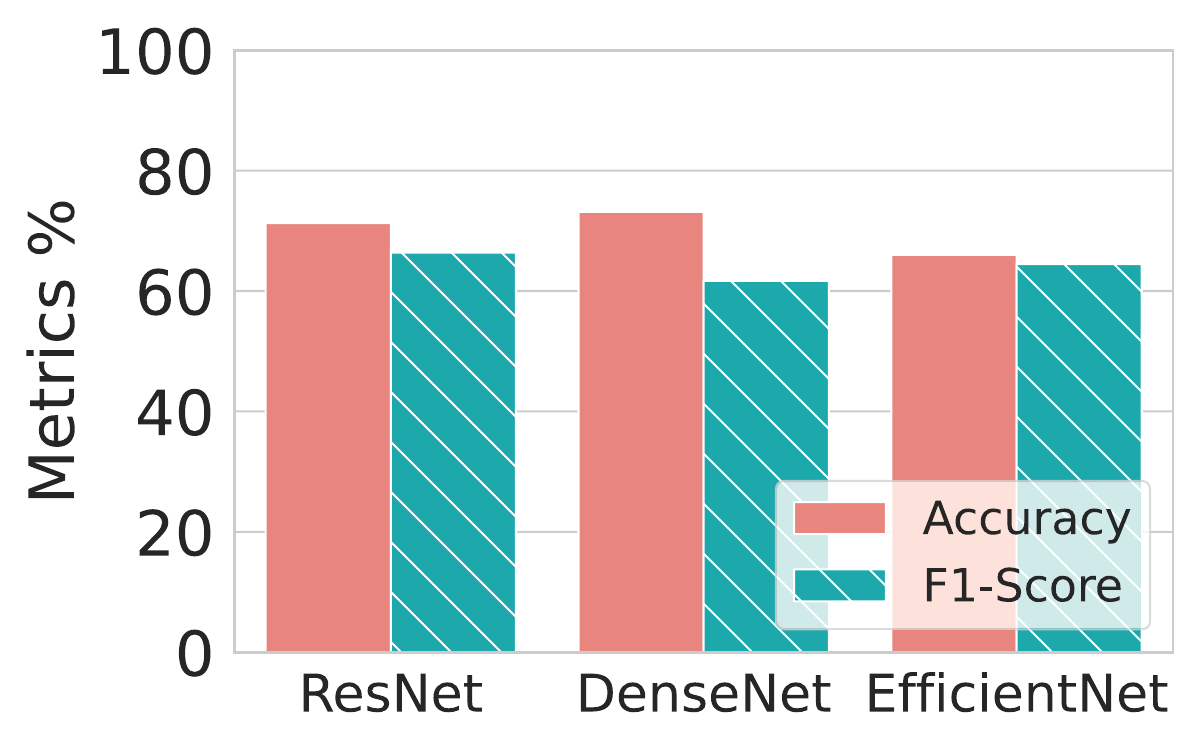}
    \label{fig:correction_imagenet}}
\vspace{-0.1in}
\caption{\sol' attack detection and re-classification models' performance across all datasets and architecture combinations. The results suggest a direct correlation between the accuracy of the \tea model and the accuracy of the detection and re-classification models.}
\label{fig:detection_correction}
\end{figure*}

We used three DNN architectures, namely ResNet, DenseNet, and EfficientNet for the image classification application. 
More specifically, we used ResNet-152 (60.4M parameters), DenseNet-201 (20.2M parameters), and EfficientNet-B7 (66.7M parameters) as the set of \tea models and ResNet-50 (25.6M parameters), DenseNet-121 (8.1M parameters), and EfficientNet-B0 (5.3M parameters) as their corresponding \stu models, respectively. 
The accuracy of the corresponding \tea and \stu models on the given datasets are tabulated in Table~\ref{tab:model_accuracy}. 
Note that the accuracy of the models we used may not reach the reported accuracy in their respective papers. However, our goal is to use them in developing a result validation mechanism for MLaaS inference.
While we used a similar architecture for each \tea and \stu model pair, it is possible to use different architectures.
%
%

For our GAN framework, we adopted the fully trained \tea model (architecture and parameter weights) as the generator model and further extended it by adding four extra fully connected layers. The additional layers take the learned decision boundary of the \tea model at the beginning of the training process and fine-tune the generator model, aiming to craft stronger attack outputs. For the attack detection and re-classification models, we used two lightweight fully connected neural networks, each containing three hidden layers containing 128, 256, and 128 neurons per hidden layer.
%
%
\subsection{System and Experiments Setup}
We evaluated our framework on a server with an Intel Xeon Platinum 8352Y processor with a base clock speed of 2.20~GHz and 256~GB RAM. The server's processor is a 3rd generation Xeon processor (\ie Ice Lake series) with 64~GB EPC. The SGX drivers are in-kernel drivers and were implemented using Gramine OS v1.2.
We used three classes of consumer processors, namely a mobile class device with a Snapdragon~765G processor, an IoT class device with an ARM Cortex-A72 processor, and an X86 consumer class device with an AMD Ryzen~5 processor.

\subsection{Verification Efficacy Analysis}
We evaluated the attack detection and re-classification model's performance in terms of accuracy and F1-score. 
In doing so, we built a test dataset by randomly selecting an average of 9000 samples from each dataset and applying the attacks we described in Section~\ref{sec:attack}. We applied the attacks either on the \tea models' outputs of the 9000 samples (for naive and advanced attacks), input samples (for adversarial example attack), or the \tea model itself (for Trojan attack). We also included the output of the legitimate \tea models on the 9000 samples to the test dataset to represent the no-attack scenario.  
In our experiments, the accuracies we reported for the re-classification models are of the detected samples. 
Note that running the standalone re-classification models on all the generated attack samples (including the undetected ones) results in higher accuracies than those we discuss below.
\begin{table}[!tb]
    \centering
    \caption{Attack detection accuracy per attack type.}
    \small
    \vspace{-0.1in}
    \begin{tabular}{lccccc}
        \hline\hline
        \multicolumn{2}{c}{}
                & & ResNet & DenseNet & EfficientNet  \\ 
        \hline
        \multirow{8}[0]{*}{\rotatebox[origin=c]{90}{CIFAR-10}}
            & \multirow{4}[0]{*}{Accuracy} 
                & Naive  & 95.07\% & 96.24\% & 97.36\% \\
                &       & Advanced  & 95.27\% & 96.36\% & 97.59\% \\
                &       & FGSM / PGD  & 89.25\% & 95.11\% & 95.89\% \\
                &       & Backdoor  & 95.23\% & 96.34\% & 97.57\% \\
            \cline{2-6}
            & \multirow{4}[0]{*}{F1-Score} 
                & Naive  & 94.84\% & 96.11\% & 97.30\% \\
                &       & Advanced  & 95.03\% & 96.22\% & 97.53\% \\
                &       & FGSM / PGD  & 89.39\% & 94.99\% & 95.86\%\\ 
                &       & Backdoor  & 94.99\% & 96.20\% & 97.53\% 
        \\\hline
        \multirow{8}[0]{*}{\rotatebox[origin=c]{90}{CIFAR-100}}
            & \multirow{4}[0]{*}{Accuracy} 
                & Naive  & 91.57\% & 89.87\% & 92.68\% \\
                &       & Advanced  & 86.62\% & 82.24\% & 90.18\% \\
                &       & FGSM / PGD  & 94.42\% & 95.63\% & 93.22\% \\
                &       & Backdoor  & 96.86\% & 97.14\% & 96.64\% \\
            \cline{2-6}
            & \multirow{4}[0]{*}{F1-Score} 
                & Naive  & 91.76\% & 90.32\% & 92.73\% \\
                &       & Advanced  & 87.52\% & 84.19\% & 90.48\% \\
                &       & FGSM / PGD  & 94.39\% & 95.58\% & 93.23\% \\
                &       & Backdoor  & 96.76\% & 97.07\% & 96.52\% 
        \\\hline
        \multirow{8}[0]{*}{\rotatebox[origin=c]{90}{ImageNet}}
            & \multirow{4}[0]{*}{Accuracy} 
                & Naive  & 84.36\% & 82.37\% & 83.56\% \\
                &       & Advanced  & 78.87\% & 76.08\% & 76.39\% \\
                &       & FGSM / PGD  & 87.94\% & 86.95\% & 85.31\% \\
                &       & Backdoor  & 89.42\% & 88.18\% & 88.22\% \\
            \cline{2-6}
            & \multirow{4}[0]{*}{F1-Score} 
                & Naive & 83.60\% & 81.40\% & 82.50\% \\
                &       & Advanced & 79.05\% & 76.34\% & 76.64\% \\
                &       & FGSM / PGD & 75.53\% & 85.53\% & 84.06\% \\
                &       & Backdoor  & 88.29\% & 86.71\% & 81.25\% \\
        \hline\hline
    \end{tabular}
    \label{tab:attack_accuracy}
    \vspace{-0.2in}
\end{table}
\begin{figure*}[!ht]
\centering
\subfigure[Memory usage of service/verification models.]{
\includegraphics[width=0.31\textwidth]{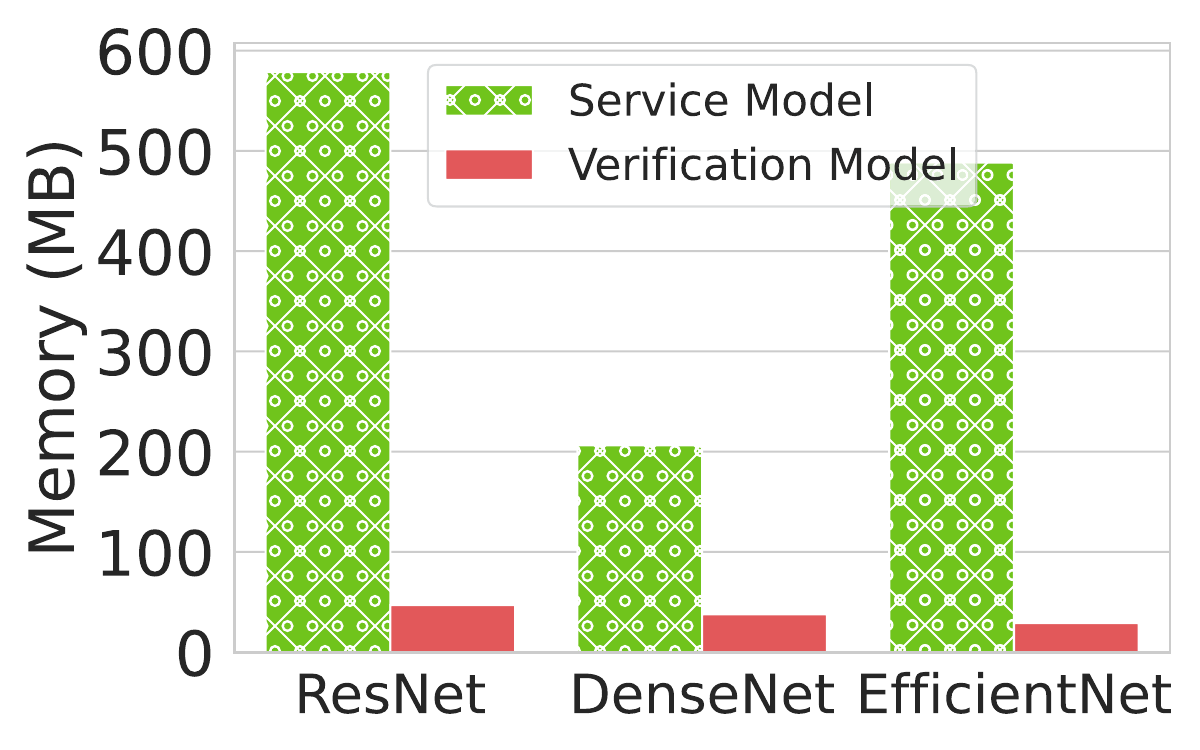}
    \label{fig:fides_performance_memory}}
\subfigure[Execution time of service/verification models.]{
\includegraphics[width=0.31\textwidth]{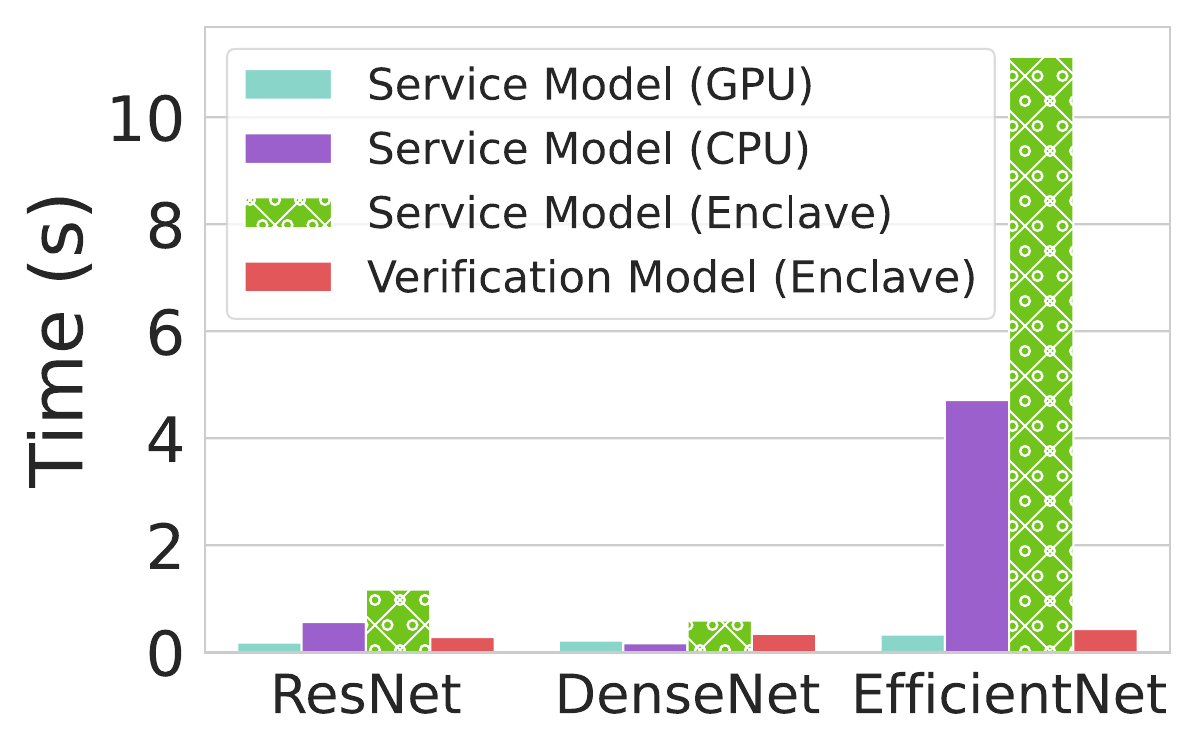}
    \label{fig:fides_performance_time}}
\subfigure[Execution time of verification techniques.]{
    \includegraphics[width=0.31\textwidth]{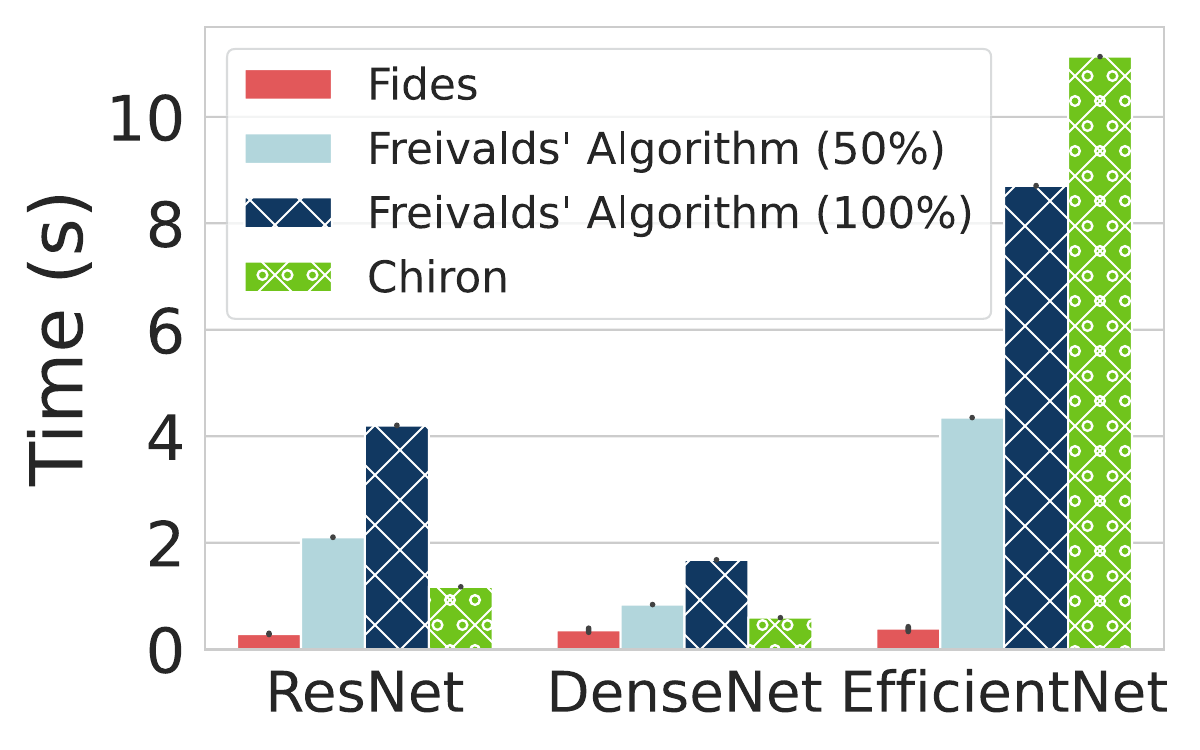}
    \label{fig:fides_comparison}}
\vspace{-0.15in}
\caption{Evaluation of \sol' server-side performance in terms of memory and execution time on the server.}
\label{fig:entroclave_time}
\vspace{-0.1in}
\end{figure*}
\begin{figure*}[!ht]
\centering
\subfigure[Inference time of attack detection models.]{
    \includegraphics[width=0.31\textwidth]{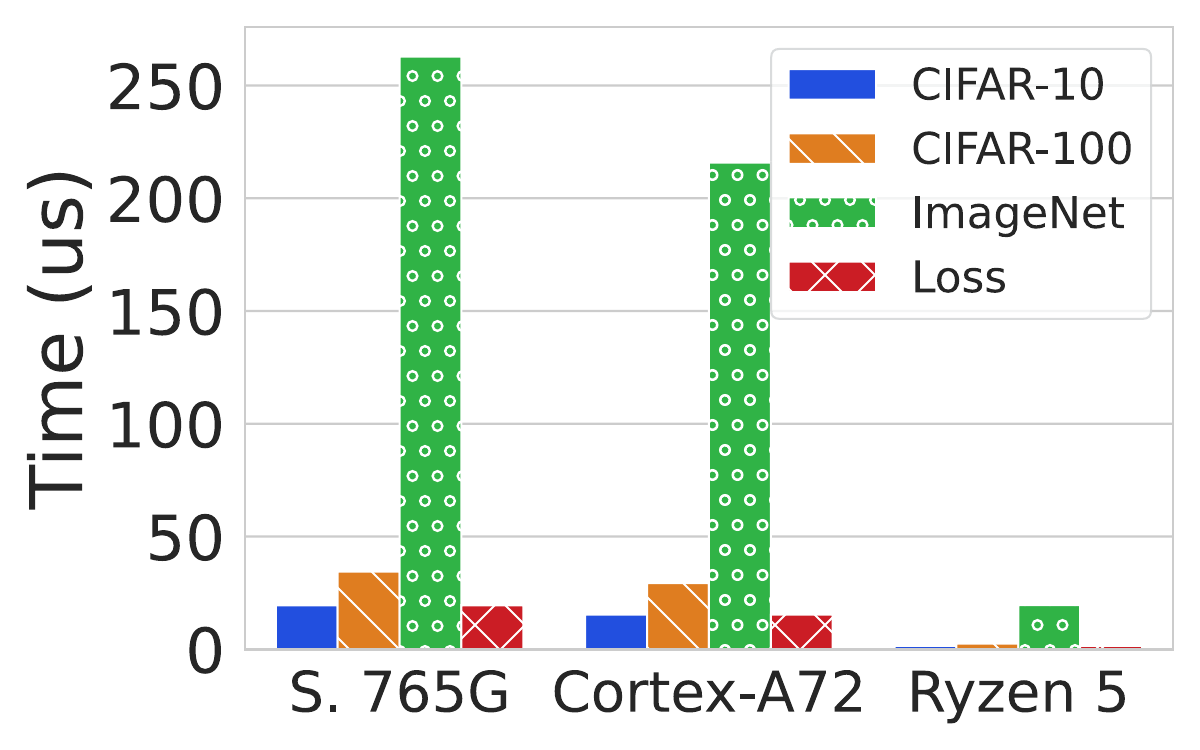}
    \label{fig:detection}}
\subfigure[Inference time of attack re-classification models.]{
    \includegraphics[width=0.31\textwidth]{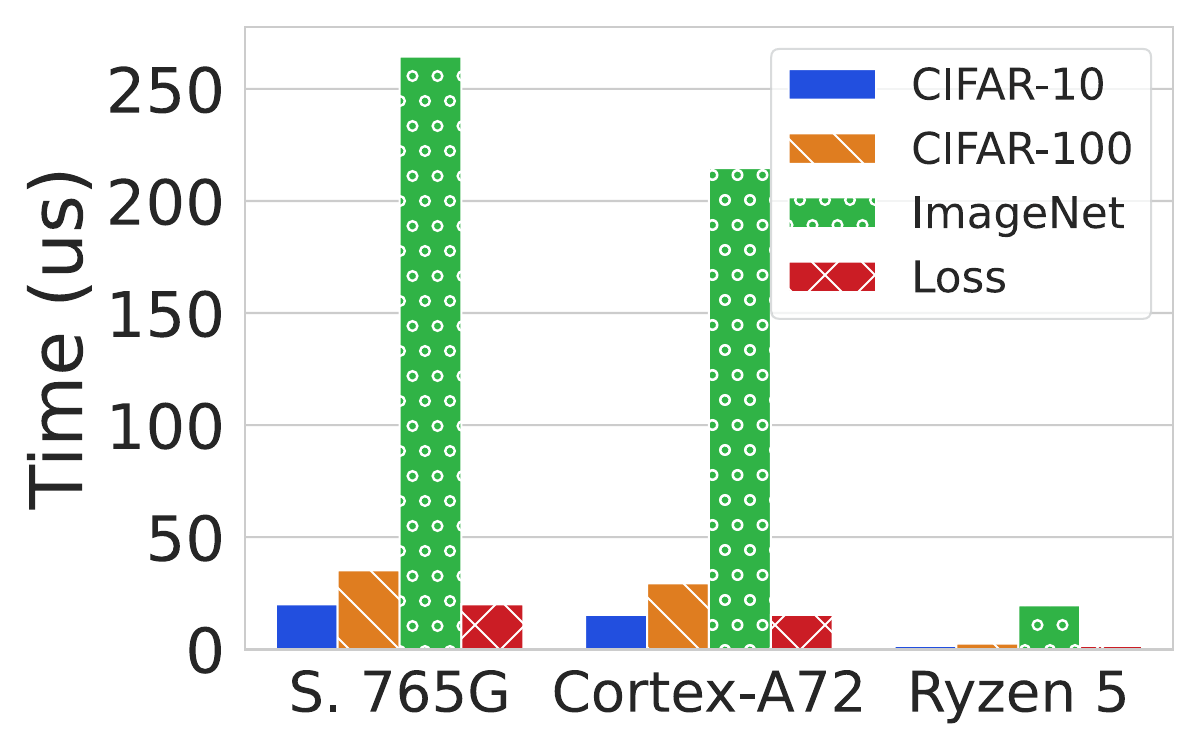}
    \label{fig:correction}}
\subfigure[Memory footprint of attack detection and re-classification models.]{
    \includegraphics[width=0.31\textwidth]{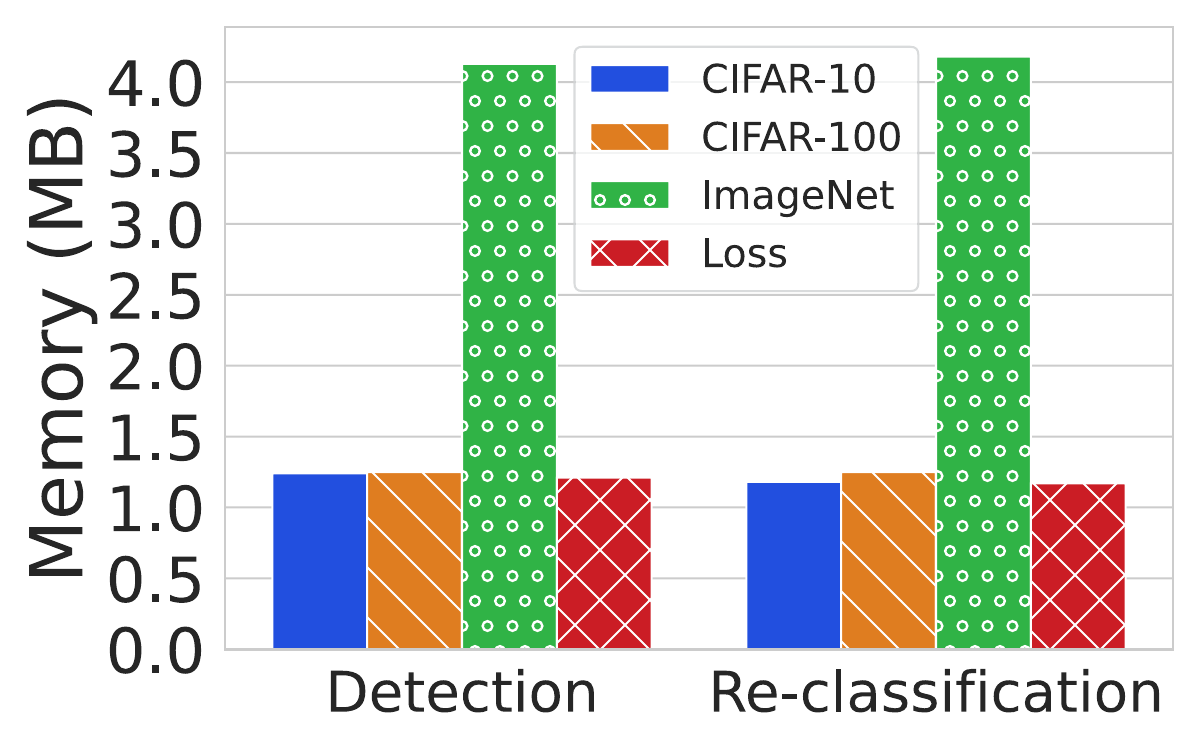}
    \label{fig:detection_correction_memory}}
\vspace{-0.15in}
\caption{The number of input features for various datasets impacts the attack detection and re-classification model's performance. Using loss (cross-entropy) for attack detection and re-classification outperforms others due to the smaller feature size.}
\label{fig:detection_correction_performance}
\vspace{-0.1in}
\end{figure*}

For the attack detection and re-classification models, we devised two variants--one that uses the soft labels of the \tea and \stu models as the input features and another one that uses the cross-entropy loss measurement as the input feature. Using the loss measurement enables the detection and re-classification models to utilize the divergence trends we discussed earlier while using soft labels allows the models to learn additional patterns, beyond the divergence trends.

Figure~\ref{fig:detection_correction} represents the accuracy and F1-Score of the loss-based attack detection and re-classification models across all architectures and datasets. We measured the attack detection accuracy across all attack scenarios, \ie from the naive to advanced and well-known attacks. Table~\ref{tab:attack_accuracy} tabulates the attack detection accuracy and F1-score for each attack type.
These results show that the attack detection models on average achieve an accuracy of 91.15\% across all architectures and datasets and the re-classification models achieve an average accuracy of 80\%. 
More specifically, using EfficientNet for CIFAR-10 results in the best attack detection and re-classification performance (Figures~\ref{fig:detection_cifar10} and~\ref{fig:correction_cifar10}). However, increasing the dataset complexity, from CIFAR-10 to ImageNet, slightly reduced the attack detection accuracy with a more moderate impact on the re-classification accuracy (Figures~\ref{fig:detection_imagenet} and~\ref{fig:correction_imagenet}). Such behavior is expected as the accuracy of the \tea and \stu models for ImageNet are lower than CIFAR-10 and CIFAR-100 accuracies across all architectures (Table~\ref{tab:model_accuracy}). 

For cases like ImageNet where the accuracy of the \tea and \stu models are (on average) 75.24\% and 70.85\%, the attack detection accuracy is still higher than 87\%, showing that the detection models were successfully identifying attacks where the \tea model and \stu model disagree (the non-trivial scenarios).
One can also observe that the performance of the detection and re-classification models are correlated to the \tea model's accuracy--the more accurate the \tea model, the more accurate the attack detection and re-classification models.
Finally, we observed that both loss-based and soft labels-based variants of the attack detection and re-classification models achieve similar performance in terms of accuracy and F1-Score.

%
%

\subsection{System Performance Analysis}
We also ran a set of experiments to assess \sol's system performance in terms of execution time and memory (Figures~\ref{fig:entroclave_time} and~\ref{fig:detection_correction_performance}). 
We used the ImageNet dataset in all architectures for assessing the memory consumption of the server-side components as ImageNet has the most number of classes--the worst-case scenario. Figure~\ref{fig:fides_performance_memory} shows that the \ddf process resulted in a smaller \stu model with a significantly smaller memory footprint. This is crucial for deploying real applications on resource-constrained edge servers. We note that the compression ratio may vary across different models depending on the architecture. Figure~\ref{fig:distillation_perf} (Section~\ref{sec:prepare}) shows the accuracy and the time taken by GDTL when compared to fine tuning (transfer learning) and distillation from scratch. GDTL performed the best accounting for both parameters.  

We compared the secure execution of the \tea and \stu models with the insecure \tea model on CPU and GPU. Figure~\ref{fig:fides_performance_time} demonstrates that the secure execution of the \stu model for ResNet and EficientNet architectures takes far less than the secure and insecure execution of their \tea models. For DenseNet, the \stu model only takes 0.178 seconds more than the insecure \tea model but it is still faster than the secure execution of the \tea model, which we attribute to the DenseNet smaller compression rate. The secure execution of the \stu models is marginally slower than GPU executions (worst case 130~milliseconds for Densenet), which is expected.

We also compared the system performance of \sol with two other techniques--Slalom~\cite{TraBon18} and Chiron~\cite{HunTylSho18}. We selected Slalom as it outperforms other solutions in terms of validation of outsourced ML inference task and selected Chiron due to its similarity with \sol in running the entire model in a TEE. For Slalom, we only implemented its verification process, which uses Freivalds' algorithm to verify the matrix multiplication operations of each linear layer. Considering the probabilistic nature of the verification process in Slalom, we benchmark the verification of 50\% and 100\% of the \tea model's matrix multiplications per architecture. We did not include the time of ECALL and OCALL operations (entering and exiting the enclave), which will result in a large latency overhead considering the frequency of these operations in Slalom.
Per Figure~\ref{fig:fides_comparison}, \sol outperforms other techniques in terms of validation time. 
It achieves a $1.73 \times$ to $25.7 \times$ speed-up compared to Chiron and a $4.8 \times$ to $26.4 \times$ speed-up compared to verification based on Freivalds' algorithm for 100\% of the matrices. This is in part due to the sequential execution of the Freivalds' algorithm in our experiment. Note that the speed-up in both cases increases with the increase in the size of the verification model. 
%

Given the constrained nature of clients' devices, we evaluated the performance of the attack detection and re-classification models on several consumer-grade device classes. Figures~\ref{fig:detection} and~\ref{fig:correction} show the elapsed time of running these models on an Internet of Things device (ARM Cortex-A72), a smartphone (Snapdragon 765G), and a personal desktop (AMD Ryzen 5). One can observe the negligible overhead of these models on clients; less than 260~microseconds in the worst case.
Our analyses show that the performance of these models is highly correlated to the size of their input features when using soft labels as the input. For instance, detecting an attack on ImageNet with 1000 classes takes (on average) $13.47 \times$ longer than detecting the same attack on CIFAR-10 with 10 classes. The same behavior can be observed from attack re-classification models. 
However, the loss-based detection and re-classification models show significant improvement in the execution time--only $0.076\%$ of the original latency in the case of ImageNet soft labels. This is primarily due to the independence of the model's input size to the number of the \tea model's classes when using cross-entropy loss value as the input feature. It is worth mentioning that the added overhead of loss calculation for the two input vectors is negligible. 
Figure~\ref{fig:detection_correction_memory} shows the memory footprint of different detection and re-classification models. The memory footprint falls within a similar range for CIFAR-10, CIFAR-100, and the loss-based solutions but increases significantly for ImageNet. It is primarily due to the smaller input layers for CIFAR-10, CIFAR-100, and loss-based solution, when compared with ImageNet's larger input layer.
%

%
%

\vspace{-0.05in}
\section{Related Work}
\label{sec:related}
%
The defenses against integrity violations in machine learning models during inference can be classified into three categories. 

{\bf Cryptographic defences} include solutions like multi-party 
computation~\cite{huang2022cheetah,sotthiwat2021partially,yuan2021practical,knott2021crypten}, proof-based systems~\cite{ghodsi2017safetynets,lee2020vcnn,madi2020computing}, various constructions of homomorphic encryption~\cite{niu2020toward,xu2020secure,madi2020computing,natarajan2021chex}.
While the primary goal of multi-party computation is protecting data privacy, this technique offers a degree of verifiability. Similarly, solutions based on homomorphic encryption provide verifiability by virtue of protecting the data. For instance, one secure inference operation of the ResNet50 model using homomorphic encryption on customized hardware takes about 970 seconds compared to only 100 milliseconds for the same operation on non-encrypted data~\cite{ReaChoKo21}. 

In contrast to these schemes, proof-based systems, \eg interactive and zero-knowledge succinct non-interactive argument of knowledge (zk-SNARK), have been extensively used for verifiable ML~\cite{ghodsi2017safetynets,lee2020vcnn,liu2021zkcnn,singh2021zero, demirel2017proof,zhang2020zero,weng2022pvcnn}. The proof-based solutions, although theoretically more representative of verification, require significant computation by the prover~\cite{lee2020vcnn} and do not scale for large ML models with many convolution layers~\cite{zhao2021veriml}. For instance, generating proof for the VGG16 model with 16 layers and a decision tree with 23 levels take 88.3 seconds and 250 seconds, respectively~\cite{liu2021zkcnn,zhang2020zero}. Moreover, the authors in~\cite{TraBon18} have shown that the best proof-based verifiable ML scheme is roughly 200 times slower when compared with running the entire model in a TEE. 

{\bf TEE-based defenses} has been used extensively in ML security and privacy, where confidentiality, privacy, and verifiability are paramount~\cite{TraBon18,gangal2020hybridtee,hashemi2021darknight,dong2022fusion}. In this domain, the main body of the literature aims at the efficient and secure execution of advanced models and large datasets in TEEs with small enclave page cache (EPC)~\cite{MoShaKat20,MoHadKat21,KumTouVij22}. These frameworks suggest partial outsourcing, in which a subset of the layers will be outsourced for secure execution. The authors in~\cite{KumTouVij22,MoShaKat20} suggested running the last few layers in one or multiple parallel enclaves and the rest of the layers by the client. Alternative approaches proposed running the linear layers in a secure enclave and the remaining layers outside in insecure memory and verifying them using Freivalds' algorithm~\cite{TraBon18}

{\bf Attack Specific Defences} could be designed to defend against backdoor attacks in the training phase~\cite{NguRieVit22, ChaSteVal17, PerGupHua20, TraLiMad18} by identifying and eliminating the updates or samples that are out-of-distribution. Since adversarial sample attacks take place in inference phase, the defenses against adversarial samples can take several different approaches, such as training discriminators to capture the difference between the training data and adversarial samples~\cite{ZuoZen21,LeeLeeLee18,EniChriWus20} or relying on invariance in the feature map and activation caused due to adversarial behavior~\cite{LuIssFor17,PerMar18,MetGenFis17}. 

\sol overcomes the shortcomings of these solutions by deploying a validation system, which relies on a significantly smaller \stu model's posterior information to validate the prediction of \tea model. \sol aims to be highly parallelizable in deployment while being adaptable to already deployed MLaaS models. \sol also aims to make the defense attack methodology agnostic.

\vspace{-0.03in}
\section{Conclusion}
\label{sec:conclusion}
%
We introduced \sol--a framework for output verification of MLaaS inference. \sol features a Greedy model distillation technique. \ddf process gradually unfreezes the layers of an off-the-shelf model for distillation-based fine-tuning. 
Along with deploying the \tea model, the provider also securely deploys the \stu model in a TEE on the server. The client then offloads the ML workload to the server by sharing its data with the \tea and \stu models. \sol also features client-side neural networks for attack detection and re-classification, which are trained via our proposed GAN framework. 
Our rigorous evaluation using multiple datasets and neural network architectures shows \sol's superiority to the existing solutions in system performance--a $1.73 \times$ to $26.4 \times$ speed-up and achieves up to 98\% attack detection and re-classification accuracy.
\vspace{-0.03in}
\section*{Acknowledgements}
This research was partially funded by the US National Science Foundation under grants \#1914635, \#2148358, and \#2133407, and the US Department of Energy grant \#DE-SC0023392. Any opinions, findings, conclusions, or recommendations expressed in this material are those of the authors and do not necessarily reflect the views of the US federal agencies.
%
\clearpage
\bibliographystyle{ACM-Reference-Format}
\bibliography{main}

\clearpage
\appendix
\section{Complementary Results}
\label{apn1} 
\begin{table}[hbt!]
    \centering
    \caption{Entropy Analysis of Jeffreys Divergence (JD) and Wasserstein Metric (WM) for cases that \tea and \stu models predictions are identical (Case A) post-attack.}
    \vspace{-0.1in}
    \begin{tabular}%
    {>{\centering\arraybackslash}p{0.005\textwidth}>{\centering\arraybackslash}p{0.01\textwidth}>{\centering\arraybackslash}p{0.07\textwidth}>{\centering\arraybackslash}p{0.1\textwidth}>{\centering\arraybackslash}p{0.1\textwidth}>{\centering\arraybackslash}p{0.1\textwidth}}
        \hline\hline
        \multicolumn{3}{c}{}
                & & Case A & \\ 
        \cline{4-6}
        \multicolumn{2}{c}{}
                & & ResNet & DenseNet & EfficientNet  \\ 
        \hline
        \multirow{8}[0]{*}{\rotatebox[origin=c]{90}{\small {CIFAR-10}}} 
            & \multirow{4}[0]{*}{WM} 
                & Pre-attack & $0.0038\pm0.0037$ & $0.0019\pm0.0018$ & $0.000013\pm0.000012$ \\
                &       & Post-attack & $2.73\pm0.76$ & $2.99\pm1.00$ & $2.98\pm0.99$ \\
            \cline{2-6}
            & \multirow{4}[0]{*}{JD} 
                & Pre-attack & $0.009\pm0.009$ & $0.004\pm0.0001$ & $0.000015\pm0.000014$ \\
                &       & Post-attack & $13.09\pm5.3$ & $13.09\pm4.37$ & $14.08\pm4.17$
        \\\hline
        \multirow{8}[0]{*}{\rotatebox[origin=c]{90}{\small {CIFAR-100}}} 
            & \multirow{4}[0]{*}{WM} 
                & Pre-attack & $0.160 \pm 0.157$ & $0.153 \pm 0.148$ &  $0.048\pm0.047$  \\
                &       & Post-attack & $26.344\pm13.344$ & $25.99\pm12.012$ & $24.11\pm12.12$ \\
            \cline{2-6}
            & \multirow{4}[0]{*}{JD} 
                & Pre-attack & $0.03\pm0.0293$ & $0.037\pm0.035$ & $0.0085\pm0.008$ \\
                &       & Post-attack & $9.53\pm3.98$ & $10.25\pm4.12$ & $9.74\pm4.07$
        \\\hline
        \multirow{8}[0]{*}{\rotatebox[origin=c]{90}{\small {ImageNet}}}
            & \multirow{4}[0]{*}{WM}
                & Pre-attack & $1.25\pm1.17$ & $2.96\pm2.64$ & $38.79\pm14.20$ \\
                &       & Post-attack & $51.96\pm46.79$ & $49.41\pm43.26$ & $85.13\pm35.97$ \\
            \cline{2-6}
            & \multirow{4}[0]{*}{JD}
                & Pre-attack & $0.035\pm0.032$ & $0.065\pm0.06$ & $0.216\pm0.092$ \\
                &       & Post-attack & $3.96\pm2.16$ & $3.31\pm1.82$ & $2.70\pm1.37$  \\
        \hline\hline
    \end{tabular}
    \label{tab:JSD_WM_A}
\end{table}

\begin{table}[!htb]
    \centering
    \caption{Entropy Analysis of Jeffreys Divergence (JD) and Wasserstein Metric (WM) for cases that \tea and \stu models predictions are different (Case B) post-attack.}
    \vspace{-0.1in}
    \begin{tabular}{lccccc}
        \hline\hline
        \multicolumn{3}{c}{}
                & & Case B &  \\ 
        \cline{4-6}
        \multicolumn{2}{c}{}
                & & ResNet & DenseNet & EfficientNet  \\ 
        \hline
        \multirow{4}[0]{*}{\rotatebox[origin=c]{90}{\footnotesize{CIFAR-10}}}
            & \multirow{2}[0]{*}{WM} 
                & Pre-attack & $1.87\pm0.63$ & $2.039\pm0.67$ & $1.83\pm0.62$\\
                &       & Post-attack & $0.95\pm0.57$ & $0.99\pm0.58$ & $0.68\pm0.47$\\
            \cline{2-6}
            & \multirow{2}[0]{*}{JD} 
                & Pre-attack & $4.76\pm1.99$ & $4.83\pm2.09$ & $4.04\pm1.55$\\
                &       & Post-attack & $1.75\pm1.36$ & $1.64\pm1.21$ & $0.873\pm0.66$
        \\\hline
        \multirow{4}[0]{*}{\rotatebox[origin=c]{90}{\scriptsize{CIFAR-100}}}
            & \multirow{2}[0]{*}{WM} 
                & Pre-attack & $23.89\pm9.648$ & $24.019 \pm 9.49$ & $22.53 \pm 9.59$\\
                &       & Post-attack & $17.11\pm9.52$ & $17.87\pm9.62$ & $16.54\pm9.25$\\
            \cline{2-6}
            & \multirow{2}[0]{*}{JD} 
                & Pre-attack & $4.77\pm1.69$ & $5.67\pm0.035$ & $4.87\pm1.90$\\
                &       & Post-attack & $3.67\pm2.523$ & $4.50\pm2.99$ & $4.12\pm3.047$
        \\\hline
        \multirow{4}[0]{*}{\rotatebox[origin=c]{90}{\footnotesize{ImageNet}}}
            & \multirow{2}[0]{*}{WM}
                & Pre-attack & $69.13\pm34.87$ & $66.13\pm34.82$ & $66.96\pm26.09$\\
                &       & Post-attack & $58.79\pm33.12$ & $55.05\pm29.94$ & $59.9\pm22.93$ \\
            \cline{2-6}
            & \multirow{2}[0]{*}{JD}
                & Pre-attack & $1.30\pm0.56$ & $1.20\pm0.50$ &  $0.94\pm0.35$\\
                &       & Post-attack & $1.07\pm0.58$ & $1.03\pm0.54$ & $0.79\pm0.34$ 
        \\\hline\hline
    \end{tabular}
    \label{tab:JSD_WM_B}
\end{table}

\end{document}